\documentclass[twocolumn,trackchanges]{aastex701}
\usepackage{tabularray}
\usepackage{float}

\begin{document}

\title{Revealing the Connection Between the Filamentary Hierarchy and Star Cluster Formation in a Simulated NGC 628 Galaxy}

\author[0009-0004-0214-1197]{Tamara R. Koletic}
\affiliation{Department of Physics and Astronomy, McMaster University, Hamilton, ON L8S 4M1}
\email[show]{koletict@mcmaster.ca}

\author[0000-0001-9956-0785]{Rachel Pillsworth}
\affiliation{Department of Physics and Astronomy, McMaster University, Hamilton, ON L8S 4M1}
\email{pillswor@mcmaster.ca}

\author[0000-0002-7605-2961]{Ralph E. Pudritz}
\affiliation{Department of Physics and Astronomy, McMaster University, Hamilton, ON L8S 4M1}
\affiliation{Origins Institute, McMaster University, Hamilton, ON L8S 4M1}
\email{pudritz@mcmaster.ca}

\begin{abstract}

There is abundant observational evidence for the hierarchical, interconnected nature of filaments in the interstellar medium (ISM) extending from galactic down to sub-parsec scales. New JWST images of NGC 628 in particular, show clusters forming along the two spiral arms of this galaxy. In this paper we investigate filament and cluster properties in an NGC 628-like multi-scale high-resolution magnetohydrodynamic simulation. We use a filament finding tool to identify filaments and derive the probability density functions (PDFs) for the filament lengths and masses. Using a clustering algorithm we identify star clusters formed between 268 to 278 Myr and follow this population as the galaxy evolves for 60 Myr, calculating their mass PDFs, average radius growth rate, and average mass loss rate. We find a power-law index of $\alpha_m = -1.35$ for the filament masses. Calculating the power-law index from our cluster mass PDF, we find a value of $\alpha_{c,m} = -1.35$ when the clusters first form, exactly our filament mass power-law index. This shows that properties of young clusters arise from the gravitational fragmentation of their host filaments.  We track the post-formation evolution of the clusters as they become unbound, increase in radius and decrease in mass yielding an ever  steeper mass power-law index. After 60 Myr, the mass power-law index is $\alpha_{c,m} = -1.55$, matching other simulations and observations.

\end{abstract}

\section{Introduction}

Observations of nearby and extragalactic structures by Physics at High Angular resolution in Nearby GalaxieS (PHANGS), James Webb Space Telescope (JWST), Herschel, Spitzer, and Hubble \citep{Temim_2024, Thilker_2023, shimajiri, andre_2022, Arzoumanian_2019, andre_2017, wang_2015, Andr__2014} show that the interstellar medium (ISM) is arranged in a hierarchy of dark lanes of gas and dust known as filaments which span scales from kpc galactic disks to the sub-pc star-forming cores \citep{hacar_initial_2022, zhao_filamentary_2024, pillsworth_filamentary_2025}. These filaments feed the scales of star formation at every step, funnelling gas into denser ridges and accumulating mass reservoirs out of which protostars can form \citep{WellsBeuther2024, HacarKonietzka2024, andre_2017, Kirk_2013}.

The growth of filaments themselves is driven by filamentary flows, where the gas accretes onto large filaments through smaller feeder filaments \citep[][Wells et al. 2025, in revision]{KumarPalmeirim2020, hacar_initial_2022}. Through high flow rates from these feeders perpendicularly onto the filaments, the gas mass inside increases. When the mass per unit length (line mass) of a filament increases beyond the critical line mass, it has reached the point where the structure has insufficient internal motions and pressure to support itself and the filament fragments. In the simplest picture, the critical line mass only depends on the thermal gas motions based on a hydrostatic cylinder of gas \citep{ostriker_1964, Inutsuka_1997, Andr__2014, pillsworth_filamentary_2025}. The critical line mass for filaments supported purely by thermal pressure can be expressed as $m_{crit, therm} \approx 2 c_s^2/G$. Furthermore, we can express this in terms of the temperatures measured and observed in the ISM, returning the approximation $m_{crit, therm} \approx 10 M_{\odot}pc^{-1} \times (T_{gas}/10 K)$ \citep{Andr__2014}. 

However, thermal gas motions are not the only contribution to the critical line mass. Non-thermal, turbulent motions in the ISM act to increase internal pressure, supporting the filament further against collapse \citep{FiegePudritz2000, SmithShetty2012, Federrath2016}. The work of \citet{zhao_filamentary_2024} showed that the turbulence spectrum in their galactic simulations follows a Burger's law down to the limit of the spatial resolution of our filaments - 25 pc.  Our simulations, which use the same setup, show that macroturbulence exists down to our filament resolution scale and is available to contribute to their support.

 The total velocity dispersion arising from gas motions will have contributions from both the sound speed and the dispersion due to non-thermal motions: $\sigma_{tot} = \sqrt{c_s^2 + \sigma_{NT}^2}$. In the absence of magnetic field contributions, we can define the total critical line mass, also known as the virial line mass \citep{FiegePudritz2000, pillsworth_filamentary_2025}. \footnote{Magnetic fields provide additional support to the filament if they have a poloidal geometry, entering into this expression as the square of the Alfv\'en speed \citep{FiegePudritz2000, zhao_filamentary_2024}}.

\[m_{crit, tot} = \frac{2\sigma_{tot}^2}{G}.\]

The structure formed through the fragmentation of filaments is dependent on the scale of the filament. Kpc scale filaments fragment to form giant molecular clouds (GMCs). Within GMCs, filaments fragment to form star clusters that will contain filaments fragmenting into individual stars \citep{zhao_filamentary_2024}. The time scale for large scale filaments to accrete enough material and become supercritical, fragmenting to form GMCs, is from 1-5 Myr \citep{zhao_filamentary_2024, Arzoumanian_2023, Hoemann_2023}. This is similar to the embedded phase of massive star formation inside these clouds of around 5 Myr, and the timescale of 1-5 Myr for the newly formed high mass stars to completely disperse the cloud \citep{Kim_2022_times, chevance_lifecycle_2019}. The lifetime of the GMC itself is about five times as long as the embedded phase.

There are multiple physical processes that produce filaments, most commonly through the collision of shock waves driven by expanding supernova shells through HII regions \citep{hacar_initial_2022, Abe_2021}. They also form due to spiral waves which are themselves a form of density shock wave \citep[][and references therein]{McKeeOstriker2007}. How these drive the formation and evolution of filaments still needs further clarification \citep{zhao_filamentary_2024}.

Star cluster formation is driven by accretion flows from the large scale ISM onto smaller filaments \citep{WellsBeuther2024,zhao_filamentary_2024, Kirk_2013}. Observations have found a strong association between the sites of star formation and filaments from JWST \citep{JWST_NGC628_Williams, RobinsonWadsley2025, Thilker_2023}. At the intersection of multiple filaments, over-dense regions form, and eventually coalesce into massive star clusters \citep{HacarKonietzka2024, Peretto2014, Galván-Madrid_2010, meyers2009, schneider1979}. An example of a massive cluster-forming region is the Serpens cloud cluster which may be continually fed by filamentary accretion flows, fuelling its ongoing evolution \citep{Kirk_2013}.

Although recent reviews on the topic have highlighted the nature of the observed hierarchy of filamentary structures \citep{hacar_initial_2022} and research has expanded said hierarchy to include the spiral arms of a galactic disk \citep{pillsworth_filamentary_2025, WilliamsSun2022, ZuckerBattersby2018}, there still remains a disconnect  between large-scale filamentary structure and star formation in the ISM. This is now changing with the advent of observations from kpc to pc scales of nearby galaxies from the PHANGS-JWST and PHANGS-ALMA surveys that have produced a robust dataset of many different star-forming galaxy morphologies \citep{PHANGS, ALMA}. These observations provide rich datasets that encourage high-resolution simulations for the ISMs and star formation in galaxies other than the Milky Way.

In this paper, we present a novel NGC 628-like galactic disk simulation and analysis of its filamentary structures and star clusters. NGC 628, also known as Messier 74 and the Phantom Galaxy, is an extensively observed `grand design' spiral galaxy \citep{JWST_NGC628}. NGC 628 is one of the best samples for studying galactic filamentary structure in strong spiral arms due to its face-on orientation and large size. The images taken by JWST reveal a detailed and complex web of gas and dust assembled into filaments, with multiple sites of star formation evident among the dense gas. We chose NGC 628 for both its identity as the JWST poster-galaxy and its clear two arm structure \citep{williams_phangs-jwst_2024}. 

In outline, \S \ref{sec:methods} of the paper describes our simulation data and methods, including the initial conditions (ICs) for our NGC 628-like galaxy simulation. In \S \ref{sec:filident} \& \ref{sec:clustident}, we explain our use of the filament finding algorithm FilFinder and clustering algorithm HDBSCAN including our handling of errors. \S \ref{sec:results} outlines our results, first focusing on the filament hierarchy properties of the galactic disk. We look at the average cluster trends and if they indicate a hereditary nature based on our filament results. Lastly, we choose six clusters that are representative of different galactic environments, studying their characteristics and how they change as the galaxy evolves. In \S \ref{sec:discussion}, we compare our results to other simulations and observational results.

\section{Data and Methods}\label{sec:methods}

\subsection{Galaxy setup}
We simulate at high-resolution, a two-arm galaxy using the \textsc{ramses} code \citep{teyssier_cosmological_2002} that includes magnetohydrodynamics, heating and cooling, and initial conditions chosen to mimic the extensively observed galaxy NGC 628. \textbf{Our NGC 628-like simulation is a modified version of the Milky Way-like setup from \citet{RobinsonWadsley2025}, which had a more massive disk than that used in the AGORA project \citep{Kim_2016}. Here we use a further revised disk mass and distribution function derived by Dr. Jerry Sellwood that is a closer match to NGC 628.} We refer the reader to \citet{zhao_filamentary_2024} for full details of the physics included in our model.

\textbf{We adopted initial conditions that best match those observed for NGC 628}, setting the gas mass to ${M}_{gas}$ $ = 1.1 \times 10^{10} \textup{ M}_{\odot}$ \citep{kahre_extinction_2018}, and the radial scale length and scale height to ${R}_{scale} = 3.1 \textup{ kpc}$ and ${z}_{scale} = 0.4 \textup{ kpc}$ respectively \citep{aniyan_resolving_2018}. Our maximum rotation speed is ${v}_{rot}= 180 \pm 9 \textup{ km/s}$ as derived from the THINGS \& HERACLES surveys \citep{aniyan_resolving_2018}. \textbf{We include an initial, old stellar population with ${v}_{rot} = 180 \textup{ km/s}$ provided to us by Dr. Jerry Sellwood.}
Our simulation includes the formation of new star particles, supernova feedback, a live dark matter halo, and magnetic fields which simulate the structure and evolution of NGC 628 in a 300 kpc box. We evolve the galaxy for 327 Myr at a resolution of 4.58 pc, saving snapshots approximately every 5 Myr. We work with data from the last 10 snapshots (which translates to approximately 60 Myr) of our simulation because it is after the first star formation epoch and approximately 100 Myr after the galaxy has reached a steady state.

\begin{figure*}
    \centering
    \includegraphics[width=7in, height=1.75in]{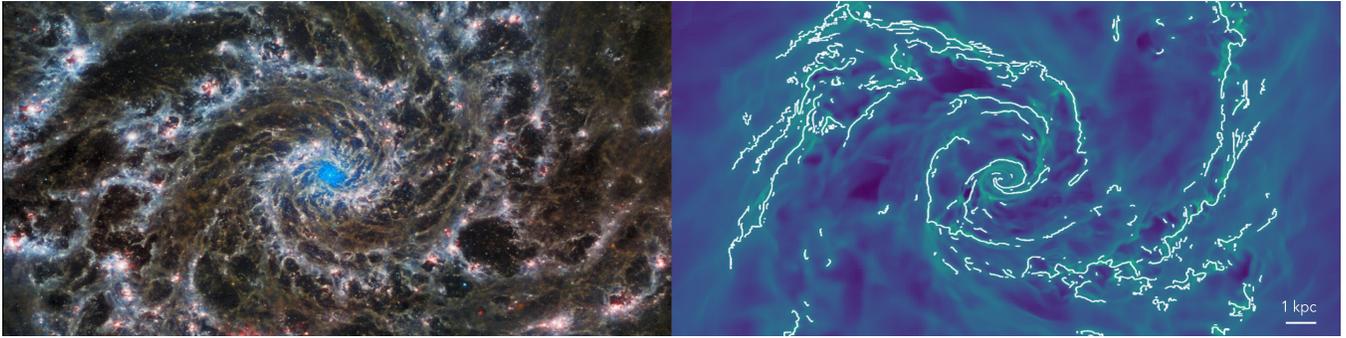}
    \caption{Left: Image of NGC 628 taken by JWST (ESA/Webb, NASA \& CSA, J. Lee and the PHANGS-JWST Team. Acknowledgement: J. Schmidt). Right: A plot of the gas density of our NGC 628-like simulation at 327 Myr zoomed-in to the center. The white lines are filaments identified using FilFinder which trace the densest gas concentrated along the spiral arms. NGC 628 simulated filament data is available on GitHub at \url{https://github.com/Tkoletic/NGC628-Filaments}.}
    \label{fig:JWST_sim_compare}
\end{figure*}

For a visual example, we show in figure \ref{fig:JWST_sim_compare} a snapshot of our simulated galactic structure next to the JWST observations of NGC 628. On the left, the observations show the center of NGC 628 in infrared, with dark voids representing the superbubbles and the dense gas and dust organized into bright white lanes. On the righthand, a column density map of the gas in our simulated NGC 628-like galaxy. We overplot the filaments, as identified with FilFinder \citep{koch_filament_2015} in white contours. We find that our simulation matches closely to NGC 628. 

\begin{figure}
    \centering
    \includegraphics[width=1\linewidth]{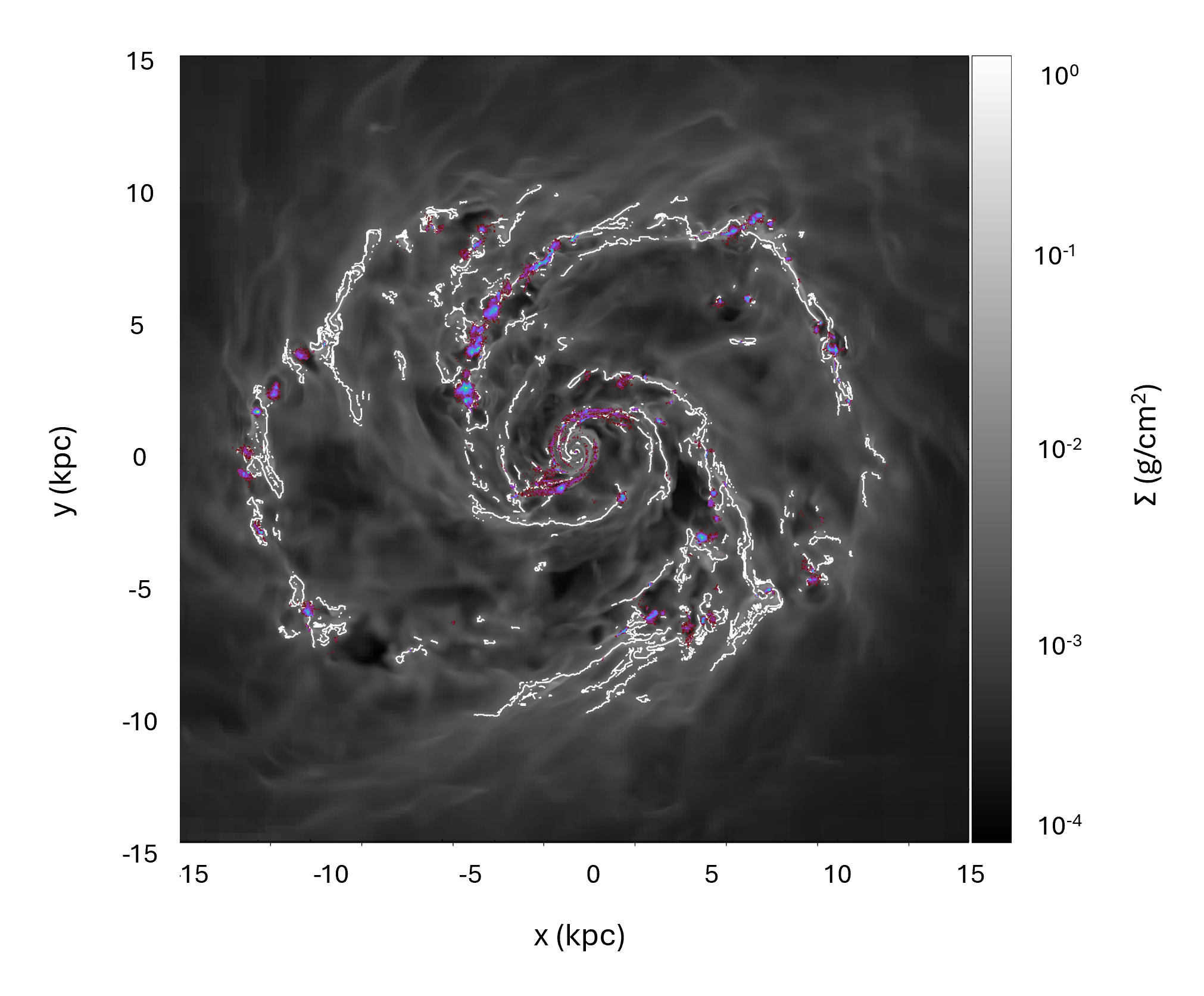}
    \caption{The gas density of our NGC 628-like simulation at 327 Myr. The white lines are filaments identified using FilFinder which trace the densest gas concentrated in the spiral arms. The clusters are of star particle formed between 307-327 Myr, semi-evenly spaced along the densest part of the arms. NGC 628 simulated filament data is available on GitHub at \url{https://github.com/Tkoletic/NGC628-Filaments}.}
    \label{fig:fil_gas_clumps}
\end{figure}

We show the full dense gas disk of our galaxy in figure \ref{fig:fil_gas_clumps}, with a greyscale column density map in the background and clusters of young star particles overplotted in the viridis map. We maintain white contours to highlight the filaments we identify. In both figures, the snapshot is taken at an age of 327 Myr.

In this simulation, we use star particles as outlined in \citet{teyssier_cosmological_2002}. The stars in our simulation have one of three masses, which vary due to the initial mass sampled from the IMF at the start of the simulation based on the resolution of their parent gas cell and the effect of mergers between star particles. However, the majority of stars formed have a mass of $198 \textup{ M}_{\odot}$, with a small fraction of star particles of masses $275 \textup{ M}_{\odot}$ and $392 \textup{ M}_{\odot}$.

We note some differences between our simulation and the real-world NGC 628. Our simulation has a lower gas column density, around $10^{-2} \textup{ g} \space \textup{cm}^{-2}$ in the midplane (figure \ref{fig:fil_gas_clumps}), compared to $2\times10^{-3} \textup{ g} \space \textup{cm}^{-2}$, the gas density of NGC 628 \citep{Chiang__2021, Sandstrom_2023}. This corresponds to a value of about $50 \space \textup{M}_\odot \space \textup{pc}^{-2} $ in our simulation and $10 \space \textup{M}_\odot \space \textup{pc}^{-2} $ from observations. However, as we are also interested in the role that galactic structure plays on the subsequent filament formation through its evolution, we have kept this column density to make comparisons between this and our Milky Way-like model more directly related to spiral arm structure. While this work is beyond the scope of this paper, it is currently in preparation.

Additionally, visual inspection reveals that our supernova feedback rates seem lower than observations of NGC 628 suggest. However, it is notable that the working frames of our simulation are still very young in the galaxy's lifetime (a maximum of 327 Myr), where feedback has not had as long to carve out multiple holes in the gas disk. We discuss in \S \ref{sec:obscomp} whether these differences make any impact on how our structures compare to the observations of NGC 628.

\subsection{Filament Identification}\label{sec:filident}
We apply the python package FilFinder, developed by \citet{koch_filament_2015}, to identify the filamentary structure in the dense gas of the galactic disk. FilFinder employs mathematical morphology to detect filaments over a large dynamic range and calculates characteristics such as width, length, and average density \citep{pillsworth_filamentary_2025}. It detects significant structures through the brightness of individual pixels in the image compared to the surrounding region. Since it can use this adaptive thresholding, FilFinder has a higher sensitivity for faint structure while also being less sensitive to bright, dense cores. From their lengths, we can determine the total mass by summing up the mass of each cell defined within the filament. 

We use a near identical FilFinder identification script to \citet{pillsworth_filamentary_2025} who showed via a parameter study that filaments are robust to variations in the masking parameters of FilFinder. We lowered the skeletonization parameter skel\_thresh from 20 to 15 pix, which acts similarly to a minimum length cut, because our galaxy is much less flocculent than \citet{pillsworth_filamentary_2025}'s and therefore less likely to misidentify structures. We set the adaptive threshold, which determines if a pixel is ``locally" bright compared to nearby pixels, to 40 pix and set the smoothing parameter to 7.5 pix, which ensures strong connectivity on typical filament width scales. A global threshold of 0.003 g$\textup{cm}{^{-2}}$ represents the global density cutoff. No smoothing is applied to the mask since we do not want an over/under estimation of our filament properties. For a more detailed explanation of filament identification using FilFinder see \citet{koch_filament_2015} and \citet{pillsworth_filamentary_2025}.

\subsection{Cluster Identification and Star Particle Selection}\label{sec:clustident}
To identify clusters we use the python clustering library Hierarchical Density-Based Spatial Clustering of Applications with Noise (HDBSCAN) developed by \citet{campello_density-based_2013}. There are many other DBSCAN variations but they are not hierarchical. We apply this algorithm to the x, y, and z coordinates of the star particles and select appropriate parameter values for the minimum cluster size and minimum samples. The minimum cluster size is the minimum number of star particles defined as a cluster; we set this to a value of 2 because this allows a minimum cluster mass of about $400 \textup{ M}_\odot$. We set a minimum number of samples to $14$, which adjusts the conservativeness of the clustering. In our determination of the best parameters, we did find within a range there is little change to the clustering. Changing the minimum cluster size to 5, for what should be the absolute minimum number in a cluster described to us through private correspondence with the authors of \citet{Claude_2023}, we find no meaningful change to our results. This makes sense because most of our clusters are so far apart it would take a drastic change in clustering parameters to merge clusters. To calculate the mass of a cluster we sum the masses of every star particle in each cluster. As a test we ran HDBSCAN using projected star particle coordinates to match the fact we use FilFinder on the projected gas density. We found this also did not meaningfully alter our results. See appendix \ref{app_c} for more information on 2D clustering.

For clustering purposes, we select star particles formed within a time frame of $10$ Myr at $268-278$ Myr. We then evolve the galaxy for $60$ Myr following the same population of stars. Thus, we have cluster data for the same population of clusters across galaxy evolution times of $10, 20, 30, 40, 50,$ and $60$ Myr. Because we do not account for velocities or the boundedness of a cluster in HDBSCAN, selecting star particles formed in a $10$ Myr period ensures by visual inspection that the star particles formed are all within clusters. Applying HDBSCAN to all star particles without using velocity data to select clusters risks star particles formed at earlier times migrating into newly formed clusters, adversely affecting the cluster distribution.

\subsection{Bayesian Analysis and Error Handling}
To determine the best fit parameters of log-normal fits, we wrote a Markov Chain Monte Carlo (MCMC) algorithm. The algorithm takes a guess for the slope and y-intercept and computes the chi-squared of the fit. It then calculates new parameters using a random variable and a defined step size. If the acceptance probability, calculated as the minimum of $1$ and $\textnormal{exp}\left( \frac{\chi^2(a_i,b_i) - \chi^2(a_t,b_t)}{2}\right)$, of the new parameters is larger than a random number between $0$ and $1$, the new parameters are accepted. The accepted parameters are added to a list. The final best-fit parameters are calculated as the average of the list of accepted values after removing the first $n$ values defined as the `burn-in'. 

It can be difficult to calculate errors in our measurements because \textsc{ramses} does not calculate uncertainty and therefore any subsequent analysis lacks error propagation. \citet{uncertainties_Kashyap} emphasize the importance of calibration errors as well as statistical uncertainty in astrophysical simulations. To calculate the uncertainty in our slope measurements, we use the numpy function quantile to compute the 69\% credible interval of our Bayesian analysis. We choose this value because it is equivalent to $2\sigma$. This returns an upper and lower bound on the best-fit parameters calculated using our MCMC algorithm.

\section{Results}\label{sec:results}

\subsection{Filament Properties}

In our simulation, the two-arm spiral appears after 100 Myr and strengthens over time. The outer ring-like shape is a product of the initial burst of star formation and begins to break apart as the density wave reaches it around 250 Myr. If we evolve the simulation further, we expect the ring to fully dissipate after 1-2 more rotations. Some of our clusters form in the dense gas in this outer ring. We have kept these in our dataset for completeness and for the sake of comparison between a true spiral arm and a purely radial structure.

\begin{figure}[h!]
    \centering
    \includegraphics[width=1\linewidth]{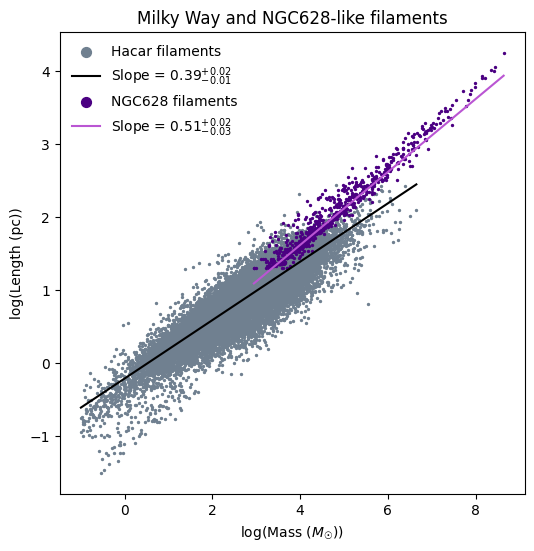}
    \caption{Log-log plot of filament length (pc) vs mass ($M_\odot$). Purple points are filaments from our NGC 628 simulation and grey are observed filaments compiled in \citet{hacar_initial_2022} and reproduced with permission from the corresponding PIs of \citet{hacar_initial_2022, SchisanoMolinari2020}. NGC 628 simulated filament data is available on GitHub at \url{https://github.com/Tkoletic/NGC628-Filaments}. The subset of Hacar filaments is publicly available \citep{hacar_initial_2022}.}
    \label{fig:hacar}
\end{figure}

As expected, the majority of the filaments form in the dense gas along the two spiral arms with few forming in the inter-arm regions. In figure \ref{fig:hacar}, we plot the lengths versus the masses of the NGC 628-like filaments, calculated from FilFinder, on a log-log plot in purple, with observed, nearby Milky Way filaments from \citet{hacar_initial_2022, SchisanoMolinari2020} in grey. We find that our NGC 628-like filaments follow a similar trend compared to the observed Milky Way filaments though there is less spread to our data. We use MCMC to calculate the best-fitting parameters to our data. In log-log space, the slope of the best fit line is $0.51^{+0.02}_{-0.03}$, resulting in a power-law trend $L \propto M^{0.51^{+0.02}_{-0.03}}$. Using the same algorithm for the Hacar data, we find a less steep power-law of $L \propto M^{0.39 ^{+0.02}_{-0.01}}$. The errors are calculated as the 68\% credible interval from the posterior distribution for each parameter \citep{whitworth2025}.

\begin{figure}[h]
    \centering
    \includegraphics[width=1\linewidth]{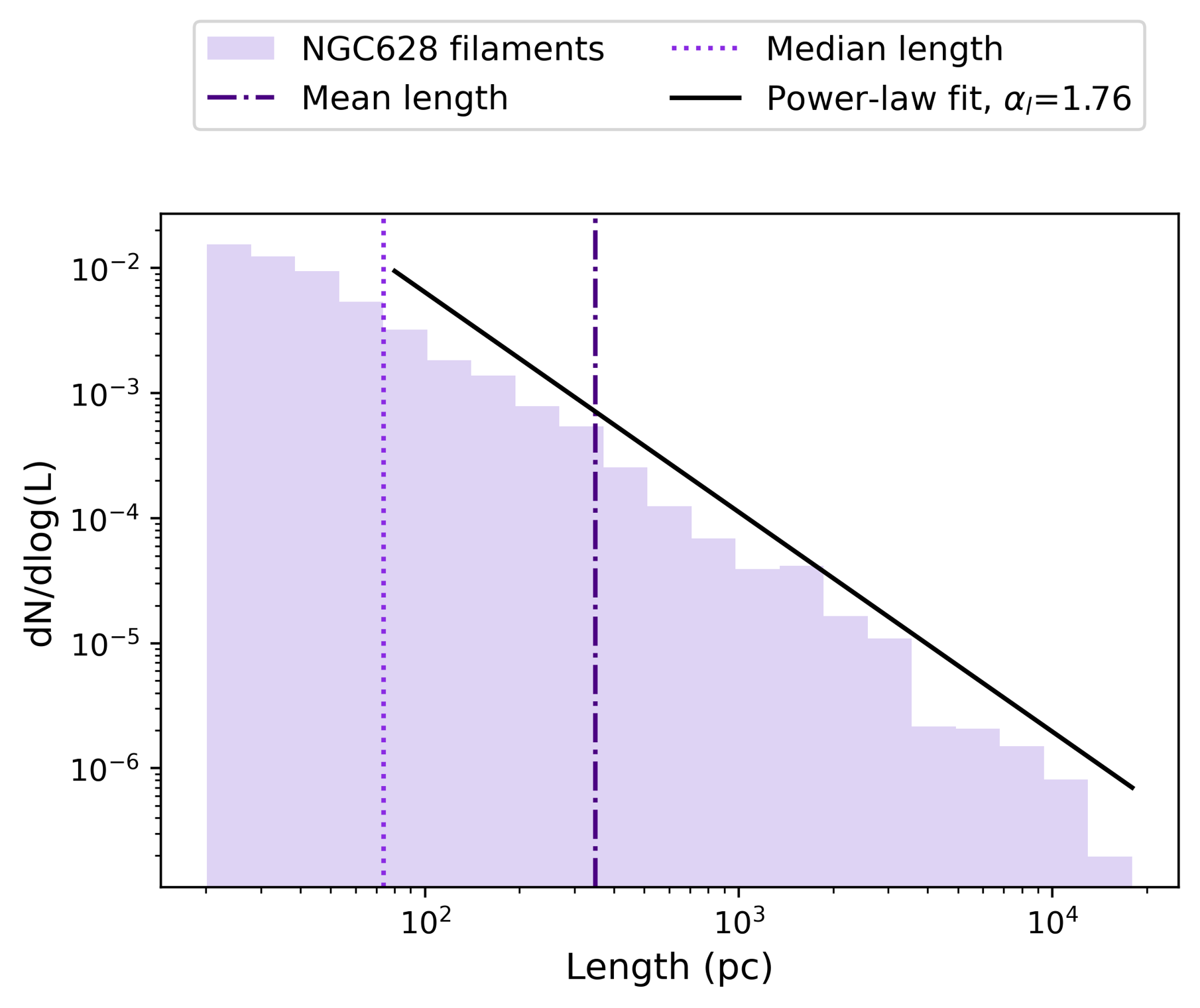}
    \caption{Length distribution of filaments. Lengths less than 25 pc were cut to avoid structures near the resolution limit of our simulation. The black line shows the power-law fit to the distribution and the dotted and dash-dotted lines indicating the median and mean filament lengths. NGC 628 simulated filament data is available on GitHub at \url{https://github.com/Tkoletic/NGC628-Filaments}.}
    \label{fig:fil_lengths_histo}
\end{figure}

Figure \ref{fig:fil_lengths_histo} shows the probability distribution function (PDF) of the lengths of the NGC 628-like filaments that span 3 orders of magnitude, with the power-law fit in black, the median filament length as the dotted line, and the mean as the dash-dotted line. Because filaments have aspect ratios of 5:1 as defined in \citet{Andr__2014} and our resolution is limited to 4.58 pc, filaments with lengths of 25 pc or shorter will have unresolved widths. Therefore, we cut all filaments shorter than 25 pc as done in \citet{pillsworth_filamentary_2025}.

The filament lengths follow a smooth distribution, indicating the presence of a structural hierarchy. Galactic filaments, on the order of a couple kiloparsecs, transition smoothly into molecular cloud scale filaments, on the order of tens of parsecs. Filaments in our simulation have an average length of 75 pc. We fit a power-law, $dN/dL \propto L^{\alpha_l}$ to our data and calculate the exponent using two different methods. 

In method one, we apply the python library, powerlaw, as used by \citet{pillsworth_filamentary_2025}, and find $\alpha_l = 1.76$. We use our separate MCMC algorithm for the second method where we feed the histogram bin centers to the algorithm to find the best-fit slope. This returns $\alpha_l = 1.74^{+0.46}_{-0.47}$. The purpose of using two methods is to compare our fits directly to \citet{pillsworth_filamentary_2025} (powerlaw library) and to determine errors on our fit (MCMC).

\begin{figure}[h]
    \centering
    \includegraphics[width=1\linewidth]{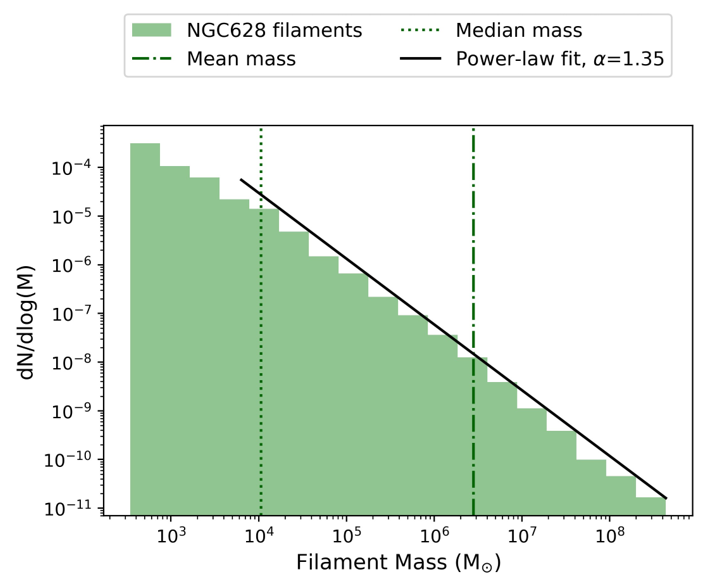}
    \caption{Mass distribution of filaments. Filaments with lengths less than 25 pc were cut to avoid structures near the resolution limit of our simulation, therefore these are also absent from the mass distribution. The black line shows the power-law fit to the distribution and the dotted and dash-dotted lines indicate the median and mean filament masses. NGC 628 simulated filament data is available on GitHub at \url{https://github.com/Tkoletic/NGC628-Filaments}.}
    \label{fig:fil_mass_histo}
\end{figure}

We employ the same process as above to statistically characterize the PDF of the filament masses for the same population of filaments (those longer than 25 pc). Figure \ref{fig:fil_mass_histo} shows the PDF with the power-law fit in black, the median mass as a dotted line, and the mean as the dash-dotted line. Applying the power-law function from python's powerlaw library, we find a fit of $\alpha_m = -1.35$. Again feeding the histogram bin centers to our MCMC algorithm, we find the best-fit slope to be $\alpha_m = -1.30^{+0.38}_{-0.39}$.

\subsection{Cluster Properties}

\begin{figure*}[ht]
    \centering
    \includegraphics[width=7in, height=4in]{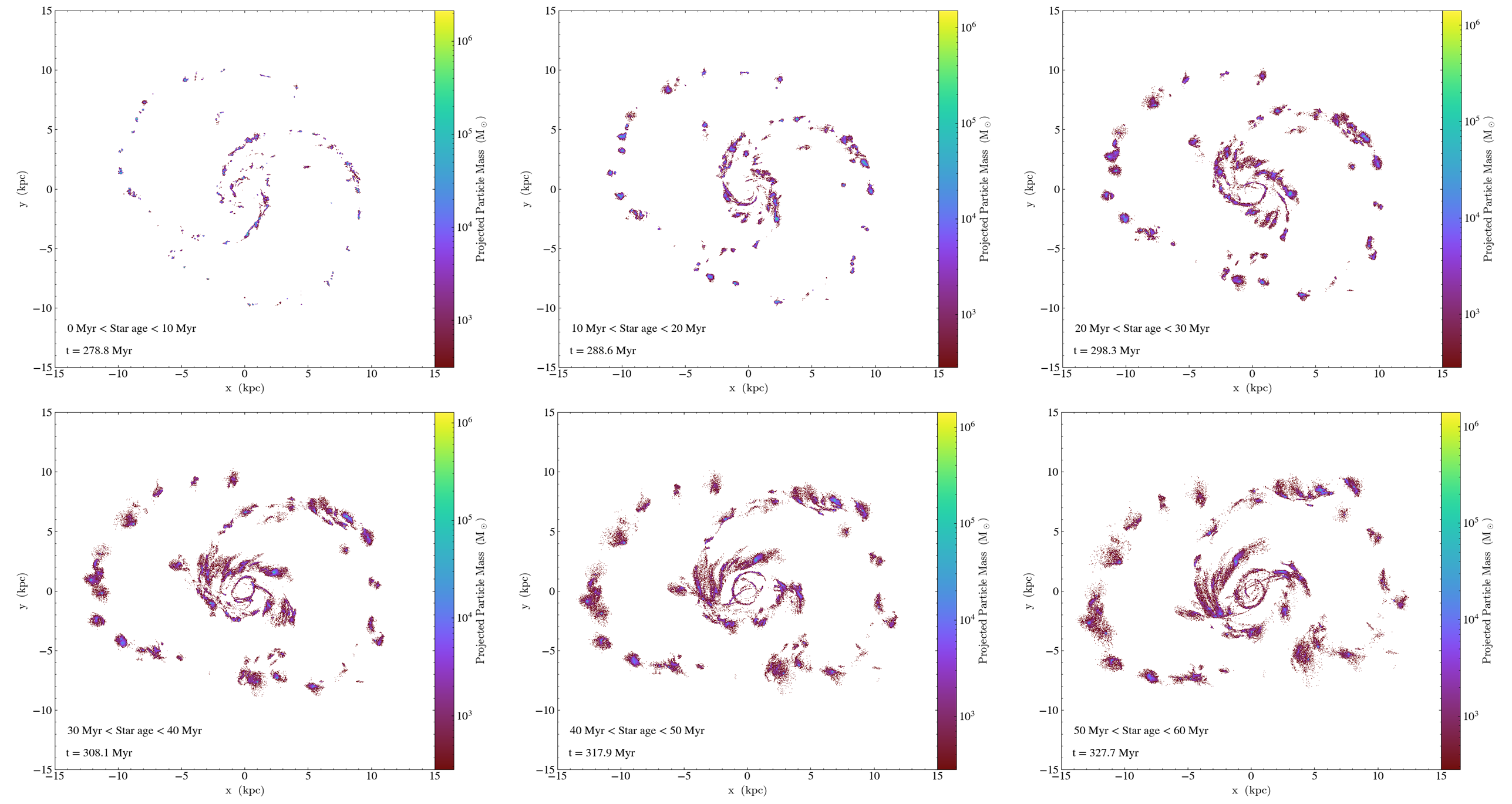}
    \caption{The first frame depicts star particle formed between $268-278$ Myr, subsequent frames show this same population after the galaxy has evolved for $ 20, 30, 40, 50,$ and $60$ Myr.}
    \label{fig:follow_clusters_ALL}
\end{figure*}

In figure \ref{fig:fil_gas_clumps}, we see new star particles forming in tight clusters along the spiral arms and along filaments. In some locations, we see a bubble blown out in the gas around the cluster, indicating a supernova explosion, while in others we see a concentration of dense gas aligned with the dense core of the cluster. If we plot only star particles formed within a $10$ Myr period, and follow this same population as the galaxy evolves, we see a number of interesting features (figure \ref{fig:follow_clusters_ALL}.) 

The stars form grouped in tight clusters, spaced 1-2 kpc apart, as evident in the first frame of figure \ref{fig:follow_clusters_ALL}. As the galaxy evolves, these stars drift outward from their cluster centers. The stars in clusters located in the outer ring appear to migrate outward isotropically while clusters in the arms, center, or inter-arm region experience strong galactic shear forces, causing the star particles to contort along the arm as the star particles are pulled towards and away from the center.

\begin{figure*}[ht]
    \centering
    \includegraphics[width=7in, height=4in]{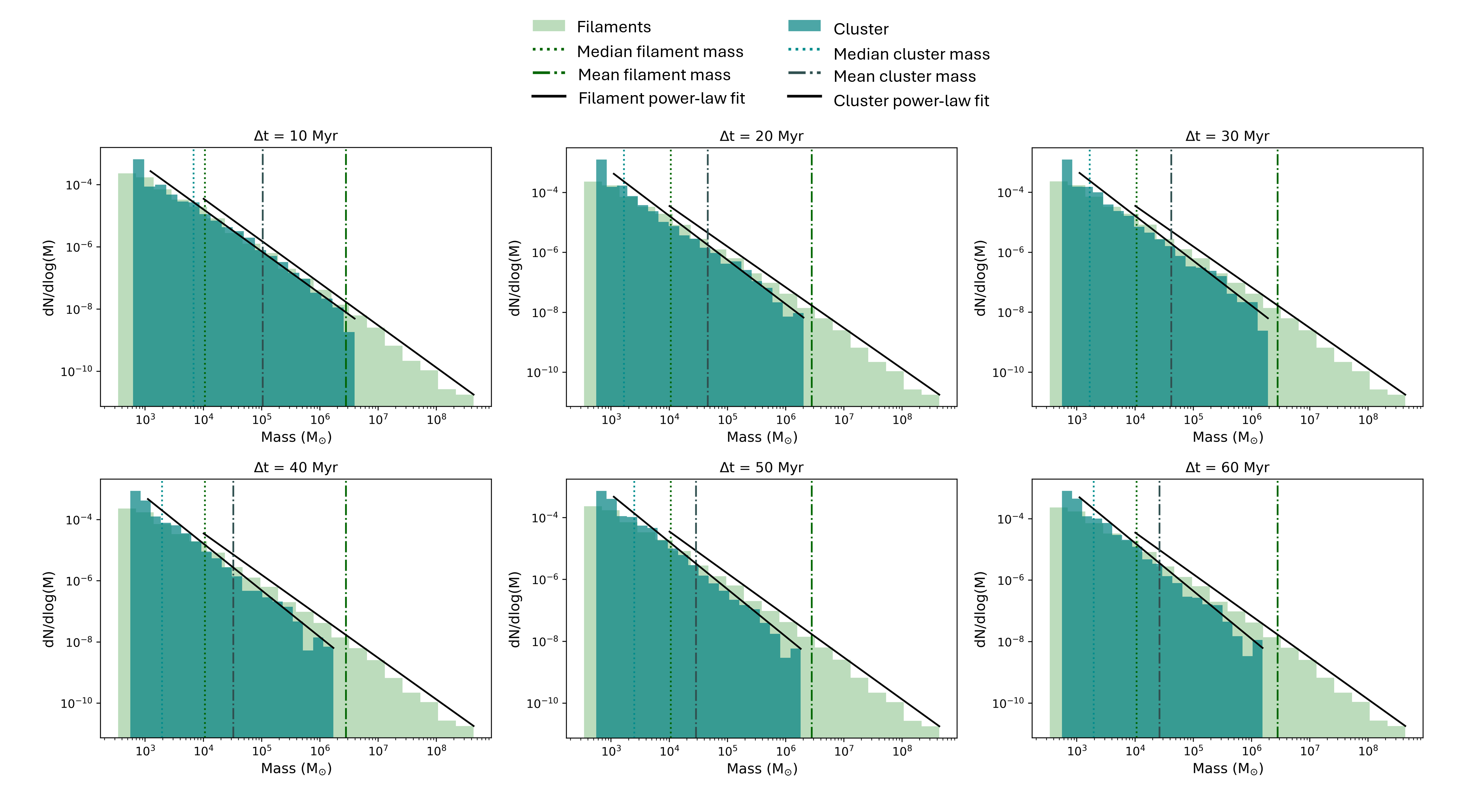}
    \caption{Six plots of the mass distribution of filaments (green) and clusters (teal) identified in 3D. The filament mass distribution is of all filaments at a time of 327 Myr. The cluster mass distribution is of the star particle population formed between 268-278 Myr. Each frame evolves the star particle population by 10 Myr, ending at 327 Myr. The black lines show the power-law fit to the filament and cluster mass distribution. The dotted and dash-dotted lines represent the median and mean masses respectively, colour coded to identify the cluster or filament distribution. }
    \label{fig:clump_fil_mass_histo}
\end{figure*}

We plot histograms of the mass PDFs of the clusters overlain on the filament mass histogram across 60 Myr in figure \ref{fig:clump_fil_mass_histo}. At formation, the mass distribution of clusters follows the exact same power-law trend ($\alpha_{c,m} = -1.35$) as the filaments and are located along dense gas in line with fragmenting peaks as shown in figure \ref{fig:fil_gas_clumps}. This provides further evidence to the theory that clusters inherit their parent filament properties \citep{zhao_filamentary_2024, pillsworth_filamentary_2025, hacar_initial_2022}. Furthermore, \citet{pillsworth_filamentary_2025} show correlations between the size of the line mass peaks of filaments and cluster sizes whereas \citet{zhao_filamentary_2024} demonstrate that clouds that overlap with filaments are in fact fragmenting out of said filaments. Since these clusters are formed out of dense clouds, it follows that if the clouds fragment out of filaments and adopt their structural properties, clusters do as well. As the galaxy evolves over 50 Myr and the star particles start to drift from their birth cluster, the power-law trend becomes steeper and the average cluster mass smaller. This is likely because as the star particles drift far enough from the cluster center, HDBSCAN no longer defines those particles as belonging to the cluster. In some cases, HDBSCAN even identifies two separate, smaller clusters that may still be one larger cluster but have drifted just far enough to be significantly separated enough for the clustering algorithm. Therefore, HDSBSCAN will preferentially identify more small clusters as the galaxy and clusters evolve. Table \ref{table_avg_cluster} gives the number of clusters of each snapshot which highlights the clustering algorithms tendency to pick up more and smaller clusters as time goes on.

\begin{table*}[ht]
\caption{Average Cluster Evolution Properties}
\label{table_avg_cluster}
\begin{tblr}{
  @{}ccccX[c,valign=b]X[c,valign=b]X[c,valign=b]@{}
}
\hline
\hline
 Evolution time (Myr)&
 Number of clusters&
 Mean mass ($M_\odot$) &
 Median mass ($M_\odot$) &
 Power-law (powerlaw library) & 
 Power-law (MCMC) &
 Mean radius (pc)\\
\hline
10 & 369 & 6818.7 & 104230.2 & 1.35 & $1.33^{+0.39}_{-0.41}$ & 2.0\\
20 & 614 & 1673.7 & 45912.0 & 1.47 & $1.39^{+0.38}_{-0.38}$ & 4.0\\
30 & 642 & 1673.7 & 42072.9 & 1.49 & $1.43^{+0.37}_{-0.38}$ &8.0\\
40 & 752 & 1952.6 & 32581.6 & 1.52 & $1.49^{+0.37}_{-0.37}$ &9.0\\
50 & 847 & 2510.5 & 28942.1 & 1.53 & $1.53^{+0.35}_{-0.36}$ &13.0\\
60 & 873 & 1952.6 & 26427.9 & 1.55 & $1.54^{+0.35}_{-0.36}$ &13.0\\
\hline
\end{tblr}
\end{table*}

\begin{figure*}[ht]
    \centering
    \includegraphics[width=7in, height=4in]{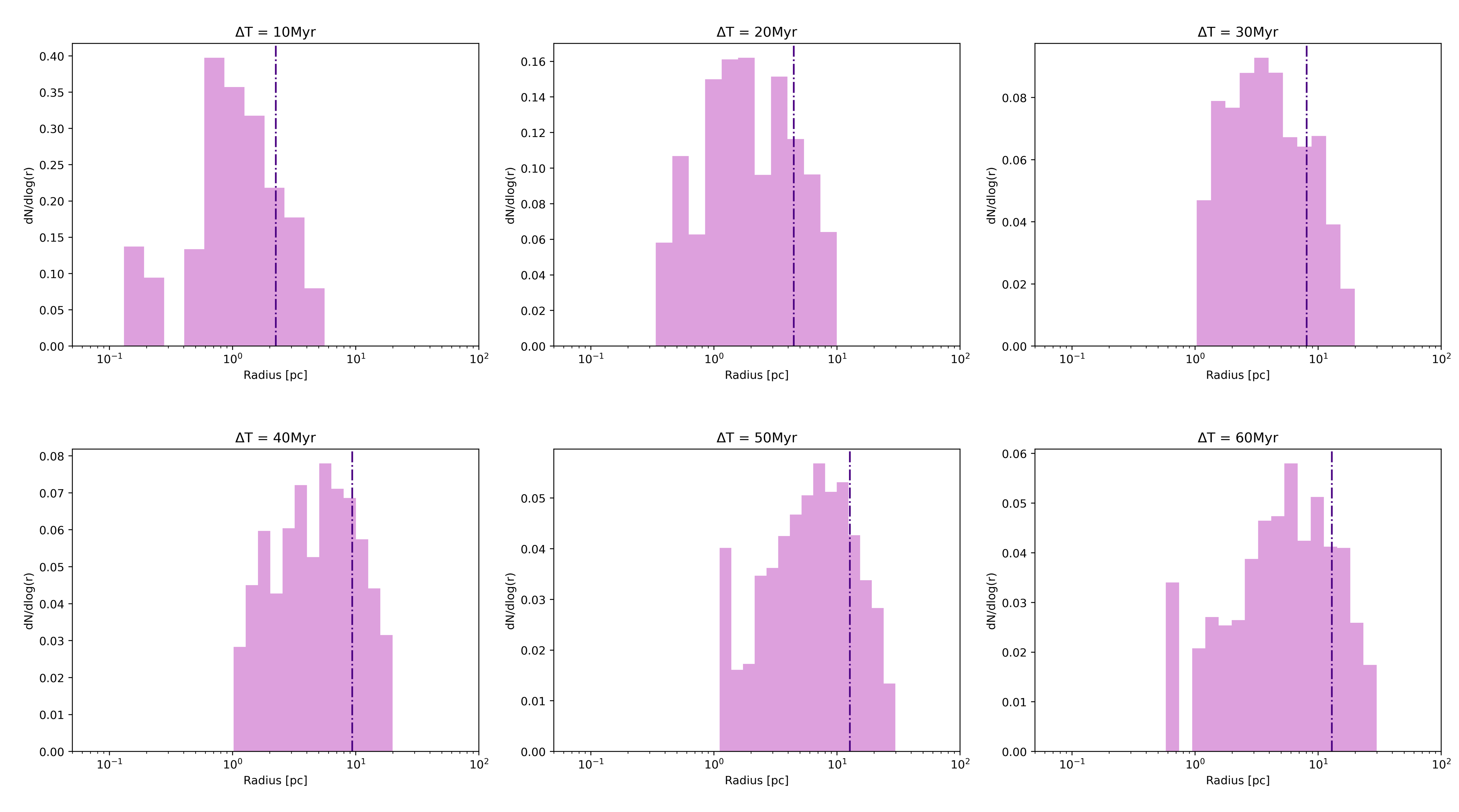}
    \caption{Six plots depicting the probability distribution function of all cluster radii at the 6 different star particle evolution times. Custom cuts were made to each time bin to remove the population of merging clusters identified by HDBSCAN as a single structure. The purple dash-dotted line is the mean cluster radius with values located in table \ref{table_avg_cluster}.}
    \label{fig:clump_radii_histos}
\end{figure*}

To observe the change in average cluster radius as the galaxy evolves, we plot histograms of the cluster radii distribution as it changes over time, shown in figure \ref{fig:clump_radii_histos}. We calculate the radii of the clusters using three different methods. The first method assumes the clusters are spherical. We average the x, y, and z positions of the particles in each cluster to determine the cluster center. We then calculate the distance between each particle and the center and generate a histogram of these distances, this is akin to placing each particle in radial bins, starting at the cluster center and moving outwards. Using the particle counts in each bin, we calculate the number and mass density profiles for each cluster. 

Our second method assumes the clusters are ellipsoidal rather than spherical, with three different axes lengths. To get an idea of the length of each axis, we take the standard deviation of the x, y, and z particle positions. We then plot a histogram of axes lengths for all clusters, as shown in figure \ref{fig:xyz_radius_histos}, where light pink is the length in z, pink the length in y, and purple the length in z. We apply this method to all clusters to determine if there is a preferential drift of particles along a single axis rather than a more isotropic radius growth assumed in the spherical method as clusters evolve and particles become unbound. The third method uses the half-mass radius as the radius of the cluster and is only used for fitting Plummer profiles to the 6 select clusters discussed in section \ref{sec:results_individual_clusters}. For all other analyses we use the first method, applied to all clusters.

Histograms of cluster radii, computed using the first method, presented with two peaks. Visually examining the shape of multiple clusters, we found some cases where HDBSCAN labelled merging or neighbouring clusters as a single cluster, with the cluster center defined between the two. Since the second peak in the histograms was much larger than a cluster, we conclude these were double clusters inaccurately identified by HDBSCAN and we cut these outliers from our data. Figure \ref{fig:clump_radii_histos} shows our final cluster radii distribution. While merging clusters are expected, with mergers contributing to half of the final cluster mass \citep{Howard_2019, karam2025dynamicsstarclusterformation}, we aim to identify an average cluster radius and trend, which will be impeded by a population of merging or neighbouring clusters. 

Overall, we see the mean cluster radius increase as the galaxy evolves. In addition to an increase in average radius overtime, there is no preferential direction of drift along any axis (x, y, or z) (figure \ref{fig:xyz_radius_histos} in appendix \ref{app_a}), demonstrating that cluster growth on average is predominantly uniform, and sees little effect from the longitudinal shear along the spiral arms of the galaxy. Another method to examine the size growth is to apply HDBSCAN to the first time step then following the exact particle IDs in a cluster. However, this would not account for mass loss. Tracking the velocities of star particles is a possible method but is out of the scope of this paper and will be addressed in future work via an in-situ clumpfinder from \textsc{ramses}.

Table \ref{table_avg_cluster} gives the mean radius, mean and median mass, and change in mass power-law for each snapshot. 

\begin{figure}[h]
    \centering
    \includegraphics[width=1\linewidth]{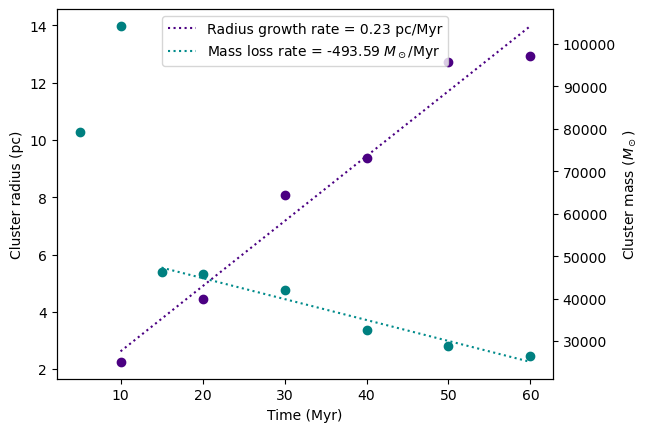}
    \caption{Radius growth rate (purple) and mass loss rate (teal) of clusters as the galaxy evolves.}
    \label{fig:clump_rate}
\end{figure}

We plot these trends in figure \ref{fig:clump_rate} and find that the radii of the clusters in our simulation grow at a rate of 0.23 pc/Myr due primarily to gas motions and galactic shear. As stars become unbound, they migrate away from the cluster and are no longer identified as belonging to the cluster. This causes a mass loss rate of $493.59 \textup{ M}_\odot \textup{ Myr}^{-1}$ after 20 Myr. We calculate the mass loss rate after 20 Myr due to a sharp increase then large drop in mass between 5 and 20 Myr. This initial increase in mass is possibly due to cluster mergers between 5 and 10 Myr \citep{Howard_2019}. We believe that the drop is due to the noise threshold we chose in the clustering algorithm. To keep the clustering consistent between time frames, we did not change the clustering parameters described in \S \ref{sec:clustident} but this meant the noise threshold was too low when particles first formed. We are unable to plot more data points between these times due to our time resolution of 5 Myr.

\begin{figure}
    \centering
    \includegraphics[width=1\linewidth]{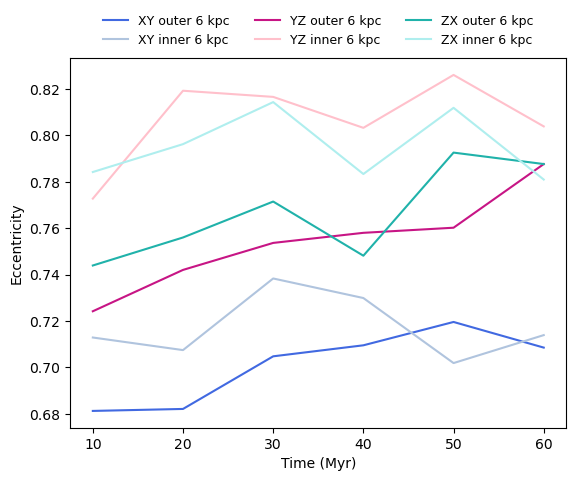}
    \caption{Average eccentricities of clusters measured along the XY, YZ and ZX axes versus time.}
    \label{fig:eccentricity_timeseries}
\end{figure}

\subsection{Selecting Individual Clusters}\label{sec:results_individual_clusters}

\begin{table*}
\centering
\caption{Select Individual Cluster Properties After 10 Myr Evolution}
\label{table_6_clust}
\begin{tblr}{
  @{}ccccccX[c,valign=b]X[c,valign=b]X[c,valign=b]@{}
}
\hline
\hline
 Cluster ID&
 Location &
 Total Mass ($M_\odot$) &
 Half-mass radius (pc) &
 Standard deviation radius (pc)\\
\hline
0 & Inter-arm & $2.75\times10^5$ & 8.72 & 6.94\\
1 & Outer ring & $9.12\times10^5$ & 12.77 & 17.78\\
2 & Outer ring & $9.13\times10^5$ & 7.76 &12.88\\
3 & Arms & $3.12\times10^5$ & 9.78 & 21.98\\
4 & Arms & $1.62\times10^5$ & 29.42 & 24.98\\
5 & Arms/Center & $3.02\times10^5$ & 28.68 & 22.90 \\
\hline
\end{tblr}
\label{table:nonlin} 
\end{table*}

\begin{figure*}[ht]
    \centering
    \includegraphics[width=5in, height=8.5in]{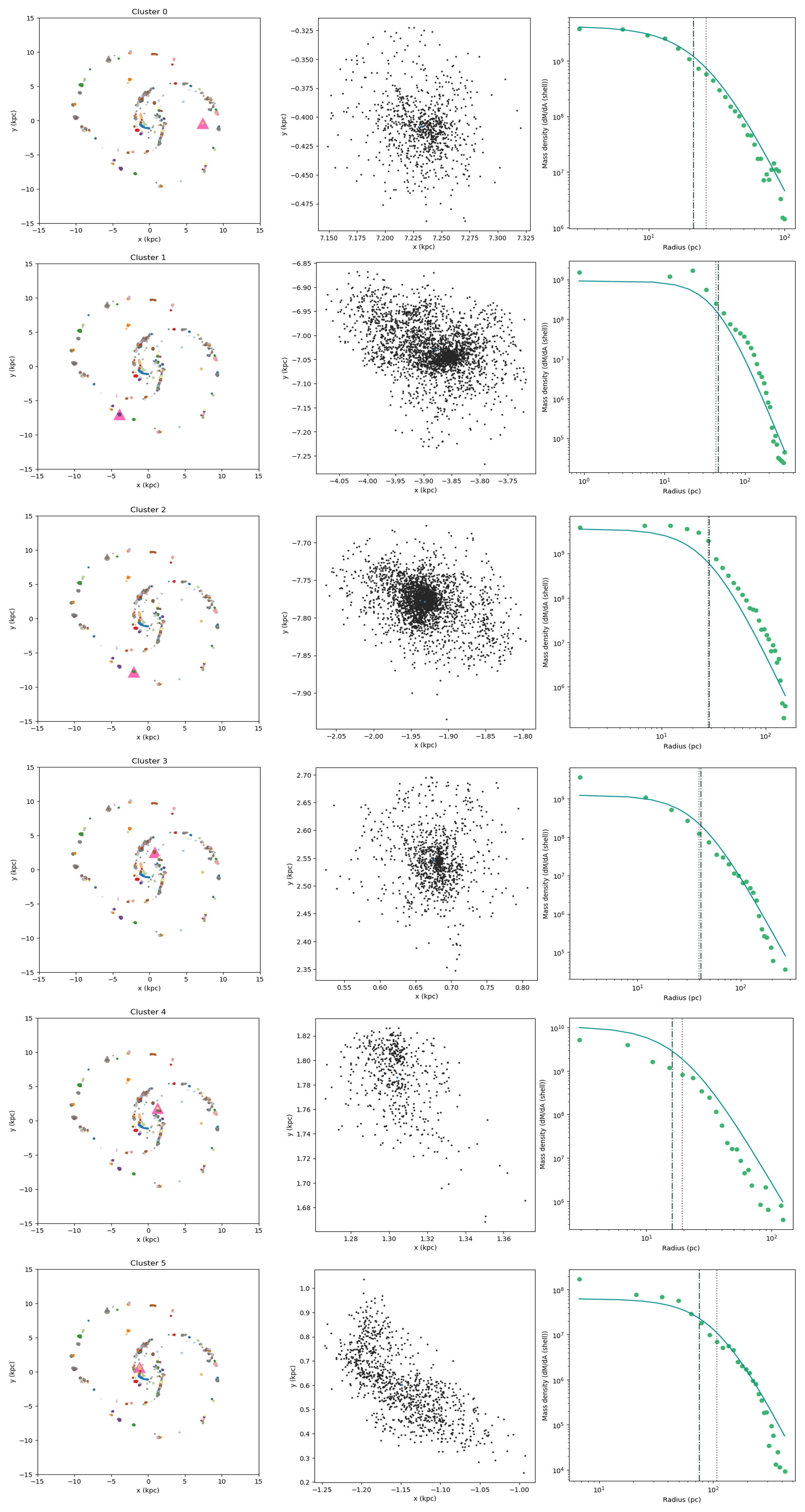}
    \caption{6 select clusters 10 Myr after the star particles were formed. The first column shows the face-on view of identified clusters where the large pink triangle represents the location of the selected cluster. The middle column shows the star particles belonging to the cluster: each point = one star particle. We identify the cluster center with a small blue star. The last column depicts the mass density profile fit with a Plummer distribution. The dash-dotted line represents the standard deviation radius and the dotted line the half-mass radius. Table \ref{table_6_clust} details the total mass and radii for each cluster.}
    \label{fig:6_select_clusters}
\end{figure*}

From the list of all identified clusters, we select six clusters to investigate their structural and evolutionary properties. We select two from the outer ring, two located in the spiral arms, one near the galactic center and one from an inter-arm region. We highlight the chosen clusters on our cluster map in figure \ref{fig:6_select_clusters}. We additionally show a zoomed-in view of the star particle distributions of these clusters and the radially binned mass density profiles of each.

Although we treat the clusters as spherical to derive an estimate of their radius, it is clear from their maps that this is an approximation. These clusters are more accurately defined as ellipsoidal, and we determine the extent of their deviation away from a sphere by calculating the average eccentricities of clusters along their three axes using the following equation:

\begin{equation}
    \sqrt{1 - \frac{b^2}{a^2}} \space\space\textup{,  where  } \space\space b>a
\end{equation}

\noindent and calculate the radius along the x, y, and z axes using the standard deviation of the particle coordinates along each axis. We then calculate three eccentricities for the XY, YZ, and ZX plane.

While our HDBSCAN parameter selection can generate clusters with particle counts as low as three, we selected high mass clusters because their particle dispersion can be better approximated as a sphere with density greatest at the cluster center. We notice that clusters formed in the arms are more elongated than those located in the outer ring. This is because clusters in the arms experience more gravitational shear. This is supported by the evolution of the eccentricities over time.

In figure \ref{fig:eccentricity_timeseries}, we show the time evolution of average cluster shape by measuring the eccentricities calculated using the x and y axis, y and z axis, and z and x axis. We split our dataset into clusters located in the outer 6 kpc and those in the inner 6 kpc of the galactic disk.
We find that the average eccentricities are higher for clusters in the inner region compared to those located in the outer ring. In future work we aim to repeat these measurements by cataloguing the drift of particles in the spiral arm's frame of reference, which may highlight the effects of shearing. 

For each selected cluster, we fit a Plummer profile of the form  

\begin{equation}
    \rho_P(r) = \frac{3M}{4\pi a^3}\left( 1+\left(\frac{r}{a}\right)^2\right)^{-\frac{5}{2}}
\end{equation}

\noindent using the half-mass radius as the Plummer radius (\cite{Plummer_prof}). Here $M$ is the mass of the cluster and $a$ is the Plummer radius. The density profiles of clusters in the outer ring are well fit by Plummer profiles since they are well approximated as a sphere due to their transport through the disk being dominated by the purely radial direction. 

\begin{figure*}
    \centering
    \includegraphics[width=7in, height=4.75in]{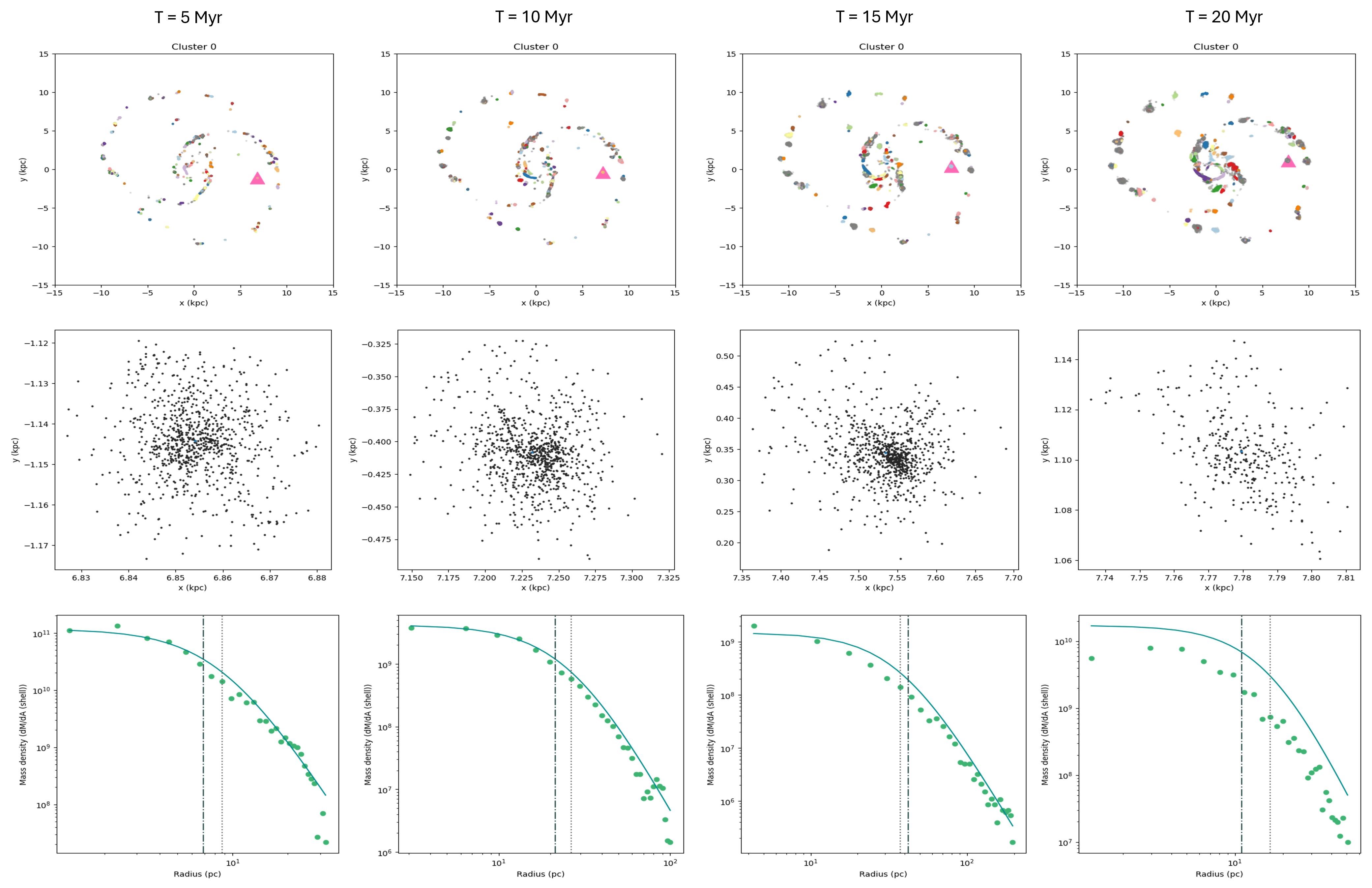}
    \caption{Cluster 0 time evolution over 20 Myr in steps of 5 Myr. The first row shows the location (pink triangle) of the cluster in the galaxy. The second row shows the scatter plot of individual star particles in the cluster. The last row shows the mass density profile where the dotted line represents the half-mass radius and the dash-dotted line is the standard deviation radius. The solid blue line is the Plummer profile where we use the half-mass radius as the Plummer radius.}
    \label{cluster_0_timeseries}
\end{figure*}

Although the selected arm and inter-arm clusters are less spherical, they too are reasonably fit by Plummer profiles. As the cluster becomes more deformed due to gravitational shear and tidal forces, it is less well fit by the Plummer profile. This is evident when following each of these six clusters as they evolve over 20 Myr, as depicted in figures \ref{cluster_0_timeseries} in this section and figures \ref{fig:cluster_1_timeseries} -  \ref{fig:cluster_5_timeseries} in Appendix \ref{app_b}. These clusters were chosen because they are centrally concentrated enough and of sufficient mass to be approximated by a sphere. Less massive clusters of only a few stellar members have very skewed density profiles and are therefore not well fit by a Plummer profile. 

\section{Discussion}\label{sec:discussion}

In this section we compare our results of filament and cluster properties to other simulations and observations. In subsection \S \ref{sec:simcomp} we compare our filament properties to those in \citet{pillsworth_filamentary_2025} and \citet{Pillsworth_2025_erratum}. We then address how our cluster radius growth and mass loss rates compare to N-body \citep{Zhou_Kroupa_2024} and isolated GC \citep{Lacchin_2021} simulations. In \S \ref{sec:obscomp} we compare our cluster mass loss rates to those measured in GCs from Gaia Data Release 3 \citep{Gaia3} and cluster mass distributions from \citet{Tang_2024}.

\subsection{Comparison to Other Simulations}\label{sec:simcomp}

We directly compare our filament properties to those presented in \citet{pillsworth_filamentary_2025, Pillsworth_2025_erratum}, since the methods are the same between those and this paper. For our filament data, we find a power-law slope of $\alpha_l = 1.74^{+0.46}_{-0.47}$, consistent with the results of $\alpha_l =1.77$ presented in \citet{pillsworth_filamentary_2025}. For our mass distributions, our fits compare just as well. While \citet{Pillsworth_2025_erratum} find a power-law slope of $\alpha_m = -1.24$ on the masses, our filament mass distribution has a slope of $\alpha_m = -1.30^{+0.38}_{-0.39}$ placing the measurement from \citet{Pillsworth_2025_erratum} in line with our own. 

These agreements among the distributions suggest that the spiral arms of a galactic disk do not play a strong role in limiting filament lengths or masses. Instead, the similarity of these results would suggest that the largest role the spiral arms play is in collecting filamentary structures into more concentrated areas -- the arms -- than a flocculent galaxy would. While the flocculent Milky Way-like model of \citet{pillsworth_filamentary_2025} has filaments throughout the galactic disk, we see our filament populations primarily gathered in the strong two-arm spiral of the galaxy, with very few inter-arm filamentary networks. 

Furthermore, since we have controlled the supernova rate and the column density of our NGC 628-like disk to be the same as in the Milky Way analogue, we can verify that the differences in the filamentary ISM's structure between the two are exclusively due to the strength of the arms. On the other hand, we would expect to see an increase in inter-arm filaments in our NGC 628-like simulation as more supernovae go off as the disk evolves. In future work, we will perform more direct comparison between the filament populations of the two disks over their evolution. 

We have also presented a cluster identification for our NGC 628-like disk, and calculated an average mass loss rate for clusters. Clusters lose mass over time through expansion, dynamical events leading to stellar ejection, and by loosely bound stars leaving the cluster due to tidal forces \citep{Laverde-Villarreal_2025}. In our simulations, we do not take into account what mass loss is from which mechanic, and instead have calculated a gross mass loss rate based on the change in the clusters members across snapshots. We find an overall mass loss rate for clusters in our NGC 628-like simulation of 493.59 $\textup{ M}_\odot \textup{ Myr}^{-1}$. 

For our primary comparison, we look at \citet{Lacchin_2021} for their N-body simulations of isolated globular clusters (GCs) with an initial mass of $10^7 M_{\odot}$ and star particle count $N = 102400$ or $25600$ with a particle mass resolution within a factor of 2 of our work. Notably, their simulations present isolated GCs, which lack information about a background galactic environment (spiral arms, supernova feedback, etc.) and do not include magnetohydrodynamics. Although we focus on star clusters within the galaxy as opposed to GCs, the initial mass of their cluster, $8.4\times10^6 \textup{ M}_{\odot}$, is the upper boundary of masses for star clusters in our simulation. 

\begin{table}
\caption{Mass Loss Fractions for Individual Clusters}
\centering
\begin{tblr}{
  @{}ccccccX[c,valign=b]X[c,valign=b]X[c,valign=b]@{}
}
\hline
\hline
 Cluster &
 Initial Mass $M_\odot$&
 Final Mass ($M_\odot$) &
 Mass Fraction\\
\hline
0 & $2.75\times10^5$ & $9.12\times10^4$ & 0.66\\
1 & $9.12\times10^5$ & $2.51\times10^5$ & 0.72\\
2 & $9.12\times10^5$ & $7.56\times10^5$ & 0.17\\
3 & $3.16\times10^5$ & $2.88\times10^5$ & 0.089\\
4 & $1.62\times10^5$ & $2.57\times10^5$ & -0.58\\
5 & $3.02\times10^5$ & $3.31\times10^4$ & 0.89\\
\hline
\end{tblr}
\label{table:mass_fraction} 
\end{table}

The fractional mass loss values presented in \citet{Lacchin_2021}'s Table 1 range from 0.44 to 0.79, depending on other parameters varied in the simulation such as particle number, primordial segregation, and depth of the central potential. We present the fractional mass loss values for each of our six example clusters in Table \ref{table:mass_fraction}. Our values range from 0.089 to 0.89, with one outlier of -0.58 which we believe underwent a merger. While most cluster mass fractions agree with those in \citet{Lacchin_2021}, some are much lower. Our large range in fractional mass loss exemplifies the difference that position in the galactic disk may play on cluster evolution. 

We calculate an average mass fraction using the median values of the initial and final masses of our six clusters. This average value is 0.75, falling in range of the \citet{Lacchin_2021} values. Thus, on average there does not appear to be a difference between the mass loss rates of young clusters in a two-arm galaxy and isolated globular clusters. In fact, if we convert the fractional values of \citet{Lacchin_2021} to a mass loss rate, we find a range of 308 $\textup{ M}_\odot \textup{ Myr}^{-1}$ to 533 $\textup{ M}_\odot \textup{ Myr}^{-1}$. With our average mass loss rate of 493.59 $\textup{ M}_\odot \textup{ Myr}^{-1}$, we see that these values also agree. Therefore, it is not just the fractional changes that are similar between different structures, but the physical mass loss rates as well. This may suggest some common drivers of mass loss between very young and very old clusters associated with a galaxy.

On top of the mass loss rates, we also find that our cluster radii are growing on average. We compare these results to N-body simulations by \citet{Zhou_Kroupa_2024} from which we estimate an expansion rate of 0.25-0.3 pc/Myr, which matches our rate of 0.23 pc/Myr and reinforce our finding that very young star clusters are in an expanding state. Furthermore, \citet{Pfalzner_2013} find that both loose and compact star clusters expand up to a factor of 10-20 within the first 20 Myr after formation, though the physical processes driving this expansion differ for their two sub-populations. Loose cluster expansion is caused by gas expulsion, whereas compact clusters expand via ejections. Although we have not separated our population of star clusters into these two groups, we do find that select clusters 2, 3, 4, and 5 expand by a factor of 10-20 in their first 20 Myr, matching \citet{Pfalzner_2013}'s comparisons between observation and simulation of cluster expansion.

Even though these four large select clusters expand by that factor, on average our clusters tend only to expand by a factor of 2 in 20 Myr, an order of magnitude lower than the value presented in \citet{Pfalzner_2013}. However, this may be due to the number of low mass clusters in our simulation. The majority of our clusters are below the $10^4 \textup{ M}_{\odot}$ cutoff in \citet{Pfalzner_2013}. Furthermore, sufficiently low-mass clusters in our simulation are represented by so few individual particles due to our mass resolution, that past a certain expansion they are no longer flagged as clusters at all. Yet, since many of our clusters with large masses match the expansion factor in \citet{Pfalzner_2013}, we can ignore the very low mass clusters in our population and find the expansion rates of the others do align with the work of \citet{Pfalzner_2013}. In future work, we will increase the mass resolution by increasing the resolution of our simulation overall and switching to sink particles, which can allow us to perform more accurate population statistics to compare to these works.

\subsection{Comparisons to Observations}\label{sec:obscomp}

Given our motivation from observation in doing this work, it is also important to compare our results to what observations tell us. Since our filament populations are so similar, we refer the reader to \citet{pillsworth_filamentary_2025} for comparison with observations on the filament side of the discussion. Instead, we focus on comparing our cluster population results to observations in the following. 

\citet{Chen_2025} calculate the mass loss rate from of 12 GCs using observations of GC streams from Gaia Data Release 3 \citep{Gaia3}. They find a mass loss range of 0.5 to 200 $\textup{M}_{\odot}$/Myr. While our measured mass loss rate does not fit this range, we note that we do not include clusters experiencing mass loss from streams. On the other hand, the wide range in values that the authors find does suggest that our measured value may not be entirely unreasonable, as they fall within the same order of magnitude as the highest mass loss rates given in \citet{Chen_2025}. Furthermore, we know that multiple mechanisms may be at work in the mass loss of our clusters. Thus, the increase in our mass loss rates compared to the observed examples may suggest that multiple mass loss drivers may work at the same time on a cluster, working to increase the mass loss rate together as opposed to one driver being the dominant change.

Most importantly, we match our cluster distributions to observed cluster mass distributions of NGC 628. \citet{Tang_2024} presents the distribution of cluster masses from 1178 observed clusters in NGC 628 using the Legacy Extra-galactic UV Survey (LEGUS). While they find a much steeper average slope of $\alpha_m \approx -2 $, Table 1 of that paper shows fit results for different populations of star clusters. Since our cluster population only contains young clusters, we can instead limit our comparison to the slope of the population omitting old, metal poor, globular clusters (OGCs) in \citet{Tang_2024}.

These OGCs are omitted because their pipeline to derive cluster demographics assumes a single, solar, metallicity. \citet{Tang_2024} notes \citet{Whitmore_2023} argue that NGC628 OGCs have much lower metallicity compared to the general population. Metallicity affects the mass function, therefore affecting the mass distribution \citep{Howard_2019}. Therefore \citet{Tang_2024} removed metal poor OGCs due to their algorithm not accurately deriving OGC masses with their assumed global metallicity. 

Without old globular clusters included, their measured slope becomes $\alpha_m =-1.50^{+0.07}_{-0.34} $. In comparison, we measure an initial slope of $\alpha_m =-1.33^{+0.39}_{-0.41}$ and a slope of $\alpha_m =-1.54^{+0.35}_{-0.36} $ after following our young cluster population for 60 Myr. Throughout the evolution of our clusters, the mass distribution matches with observed cluster mass distributions. 

Furthermore, these comparative distributions also highlight the importance of the filamentary network in NGC 628. Given the consistency between our cluster and filament mass distributions, we infer that the clusters must form from the filamentary network we identify. While \citet{Tang_2024} do not discuss the filamentary network in NGC 628, the match between our simulated clusters and their observed ones highlights that a connection between filaments and clusters very likely also exists in the actual NGC 628. As such, the filamentary hierarchy and its subsequent structure formation play an important role in the star formation of a galaxy and may give crucial context to observed cluster properties. 

\section{Conclusions}

In this paper we have simulated a strong, two-arm spiral with initial conditions matching observations of NGC 628, a galaxy extensively observed by JWST and PHANGS. We produce spiral arms within 100 Myr of the galaxy's evolution. An outer ring-like structure appears as a consequence of starting with a smooth gas disk in our ICs, causing a burst of star formation at the beginning of our simulation. While the star formation rate steadies out fairly quickly \citep{RobinsonWadsley2025}, the outer ring structure takes longer to propagate out of the galactic disk or break up into the spiral arms. 

We used the filament finding tool, FilFinder, to identify filaments in our NGC 628-like simulation and characterize the power-law fits to both the length and mass distributions. We find filaments ranging from 25 pc to 10s of kpc, building out the upper length scale of the filamentary ISM and matching with previous work on a Milky Way-like galaxy \citep{pillsworth_filamentary_2025}. Between our NGC 628-like and the Milky Way model of \citet{pillsworth_filamentary_2025}, filament distributions are the same in both length and mass suggesting that the spiral arms of a galaxy do not play a significant role in the morphology of filaments despite the clear role they play in their locations. We identified young clusters in our simulation using HDBSCAN \citep{campello_density-based_2013} and follow their evolution in the galaxy for 60 Myr. We compare their properties to both observations and simulations of clusters. Comparing to our filament mass distributions, we find that when clusters first form they follow an identical mass distribution trend as the filaments and gradually shift as they evolve and may be less influenced by the surrounding gas. 

Our specific conclusions are as follows.

\begin{enumerate}

  \item Plotting the mass vs length of filaments in a log-log plot, we find the filaments follow a trend of $L \propto M^{0.51^{+0.02}_{-0.03}}$. Including data-points of nearby filaments in the Milky Way from \citet{hacar_initial_2022}, the lower end of our filaments overlap with the largest observed Hacar filaments. The Hacar filaments follow a less-steep power-law trend $L \propto M^{0.39 ^{+0.02}_{-0.01}}$.
  
  \item The probability distribution function (PDF) of the filament lengths for lengths $\geq$25 pc follows a power-law distribution with $\alpha_l = 1.76$, matching that of the Milky Way-like simulation from \citet{Pillsworth_2025_erratum} which found $\alpha_l = 1.77$.
  
  \item Similarly, the PDF of the filament masses also follows a power-law trend. Calculating the index using our MCMC algorithm rather than the python powerlaw library used in \citet{pillsworth_filamentary_2025}, we find $\alpha_m = -1.30^{+0.38}_{-0.39}$. This is consistent with both the filaments from \citet{Pillsworth_2025_erratum} and the Milky Way cloud catalogue of \citet{Rice}. 

  \item Our cluster masses follow the same power-law trend when star particles first form in clusters, with an $\alpha_{c,m} = -1.35$. This provides evidence to the theory star clusters inherit their host filament properties. Furthermore, our power-law index after the star clusters have evolved for 60 Myr matches the value of $\alpha_m =-1.50^{+0.07}_{-0.34} $ from \citet{Tang_2024} for observed clusters in NCG 628, neglecting old GCs.

  \item Our clusters expand at a rate of 0.23 pc/Myr. Our sufficiently massive clusters have expansion rates matching those presented in \citet{Pfalzner_2013}. Moreover, our clusters experience mass loss rates of 493.59 $M_\odot/$Myr, in line with the clusters simulated in \citet{Lacchin_2021}. 

  \item The mass loss rates of our clusters are higher, but not considerably more so than the observed mass loss rate of GCs in streams from \citet{Chen_2025}. This suggests that in the presence of multiple mass loss drivers, the effects will add together to increase mass loss, as opposed to being dominated by one driver. 
  
  \item We see no preference in the particle drift in clusters along any specific axis, nor do we find significant changes in eccentricities over their evolution, suggesting that shearing motions of spiral arms in a galactic disk do not play a significant role in the expansion of clusters in their first 60 Myr. 
  
  \item However, we do find a 4-8\% difference in the eccentricities of clusters from the arms to our IC-driven ``outer ring", which may suggest that spiral arms play their role in the initial formation of clusters as opposed to throughout their evolution. Future work will investigate the evolution of clusters in situ in both the star particles and the gas to further investigate these effects.
  
\end{enumerate}

\begin{acknowledgments}
The authors thank the anonymous referee for their useful referee report. We also thank Dr. Jerry Sellwood for generating initial particle files for an NGC 628-like galaxy; Dr. Romain Teyssier, author of \textsc{ramses}, for invaluable discussions about the code; Drs. Hector Robinson and James Wadsley for their modifications to the \textsc{ramses} source code for our specific galaxy simulation; and Dr. Eric Koch for his support with FilFinder. The computational resources that made this project possible are due to a grant to REP from the Digital Research Alliance of Canada for the Cedar computing cluster. RP acknowledges funding support from an NSERC Canada Graduate Scholarship - Doctoral. REP acknowledges the generous funding support of an NSERC Discovery Grant. The data from this work is available upon reasonable request to the corresponding author. 

\end{acknowledgments}

\software{matplotlib \citep{matplotlib}, astropy \citep{astropy_2013, astropy_2018, astropy_2022}, scipy \citep{scipy}, pandas \citep{pandas, the_pandas_development_team_2025_17229934}, powerlaw \citep{powerlaw}, numpy \citep{numpy}, FilFinder \citep{koch_filament_2015}, HDBSCAN \citep{campello_density-based_2013}}

\bibliography{paper1}{}
\bibliographystyle{aasjournalv7}

\appendix\label{sec:Appendix}

\section{Evolution of cluster radii by axis}\label{app_a}
\begin{figure*}[h]
    \centering
    \includegraphics[width=7in, height=4in]{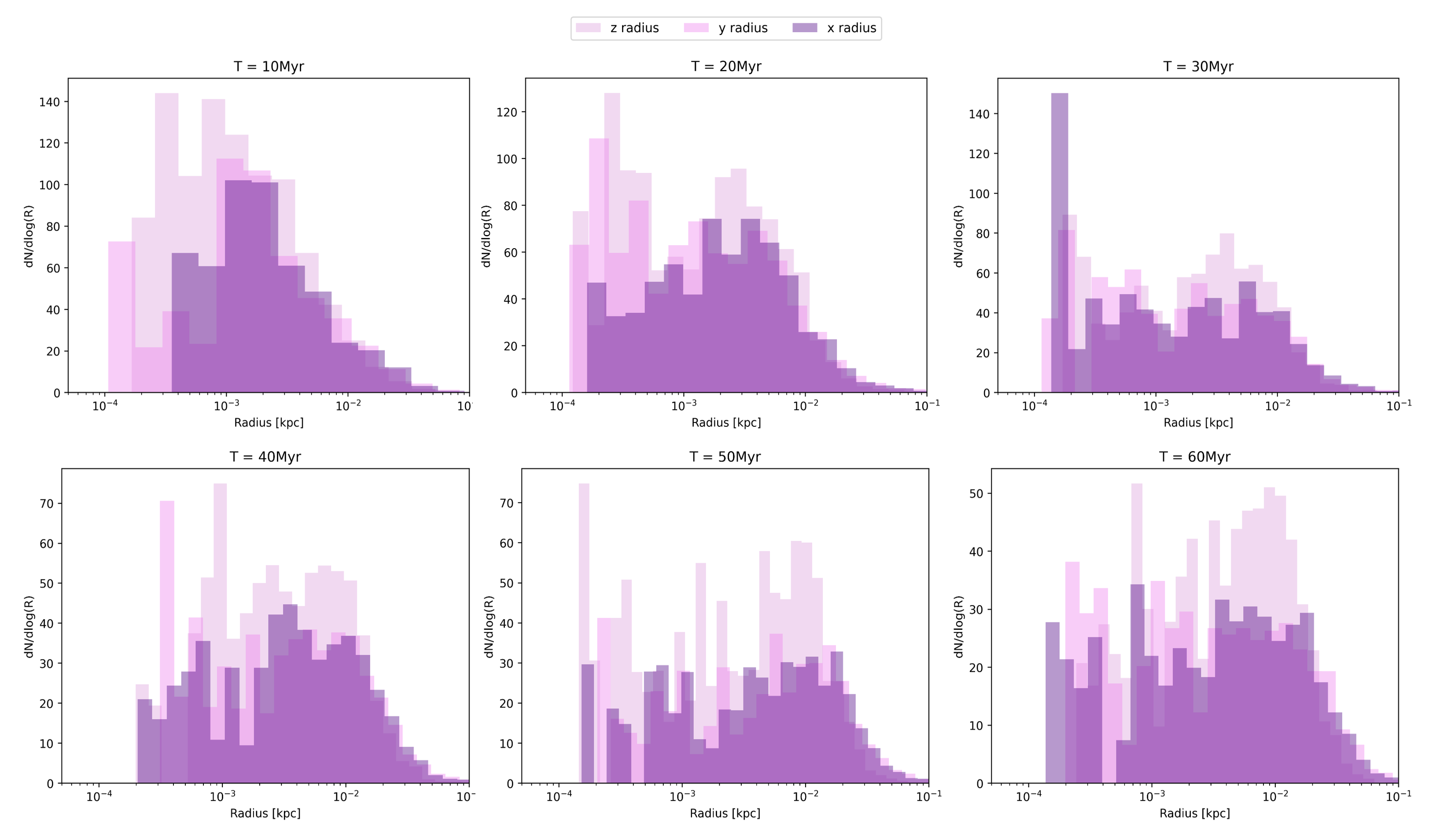}
    \caption{Radius histograms of the standard deviation x, y, and z axis lengths of clusters in steps of 10 Myr. This shows there is no preferential star particle drift along a single axis.}
    \label{fig:xyz_radius_histos}
\end{figure*}

Figure \ref{fig:xyz_radius_histos} shows six histograms in steps of 10 Myr of the distribution in the lengths of clusters measured along the x, y, and z axes. We see no shift along a single axis meaning that clusters increase in radius approximately isotropically. 

\section{Evolution of cluster properties}\label{app_b}

\begin{figure*}
    \centering
    \includegraphics[width=6.5in, height=4.0in]{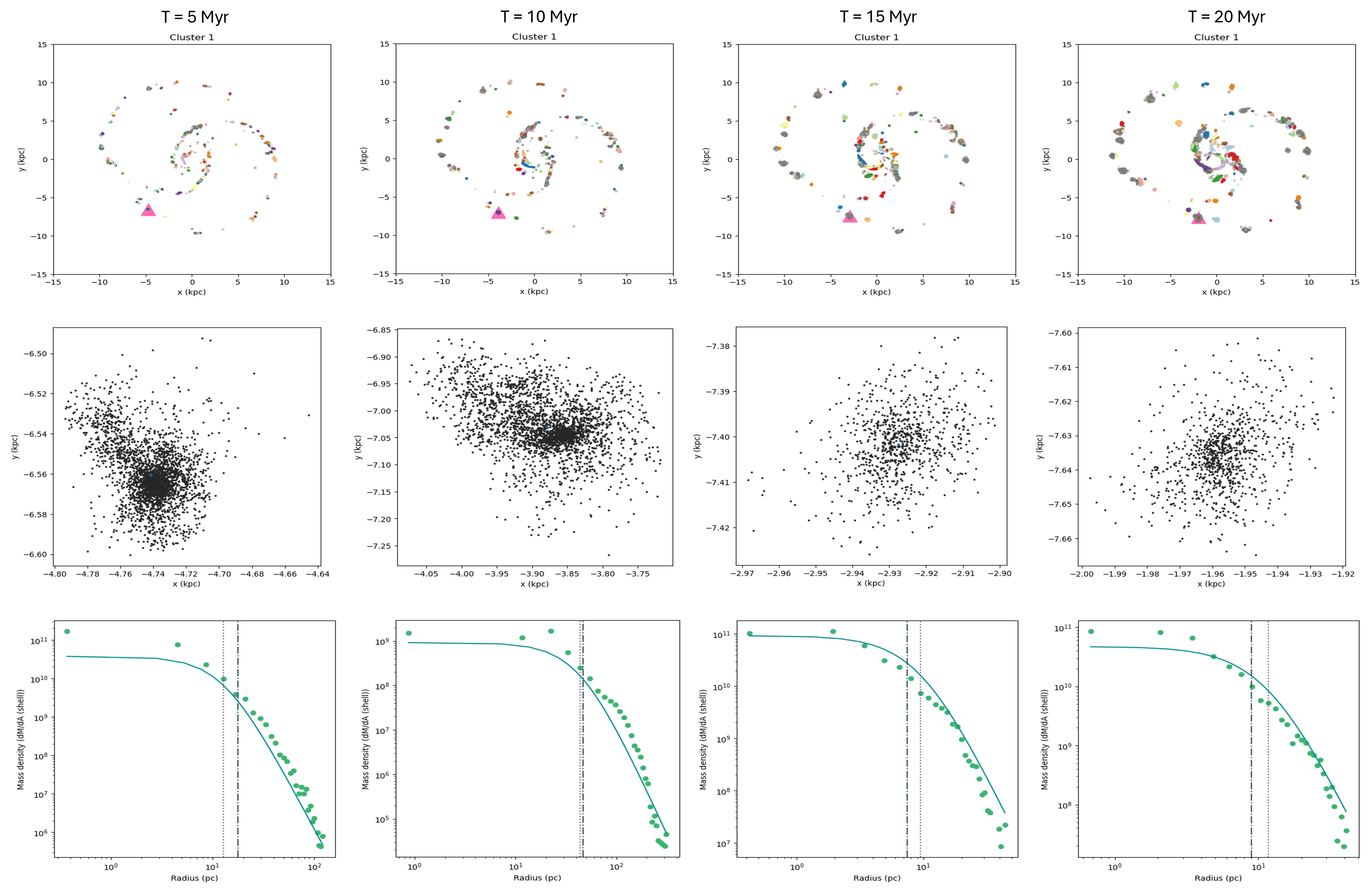}
    \caption{Cluster 1. Description of plots is the same as figure \ref{cluster_0_timeseries}.}
    \label{fig:cluster_1_timeseries}
\end{figure*}

\begin{figure*}
    \centering
    \includegraphics[width=6.5in, height=4.0in]{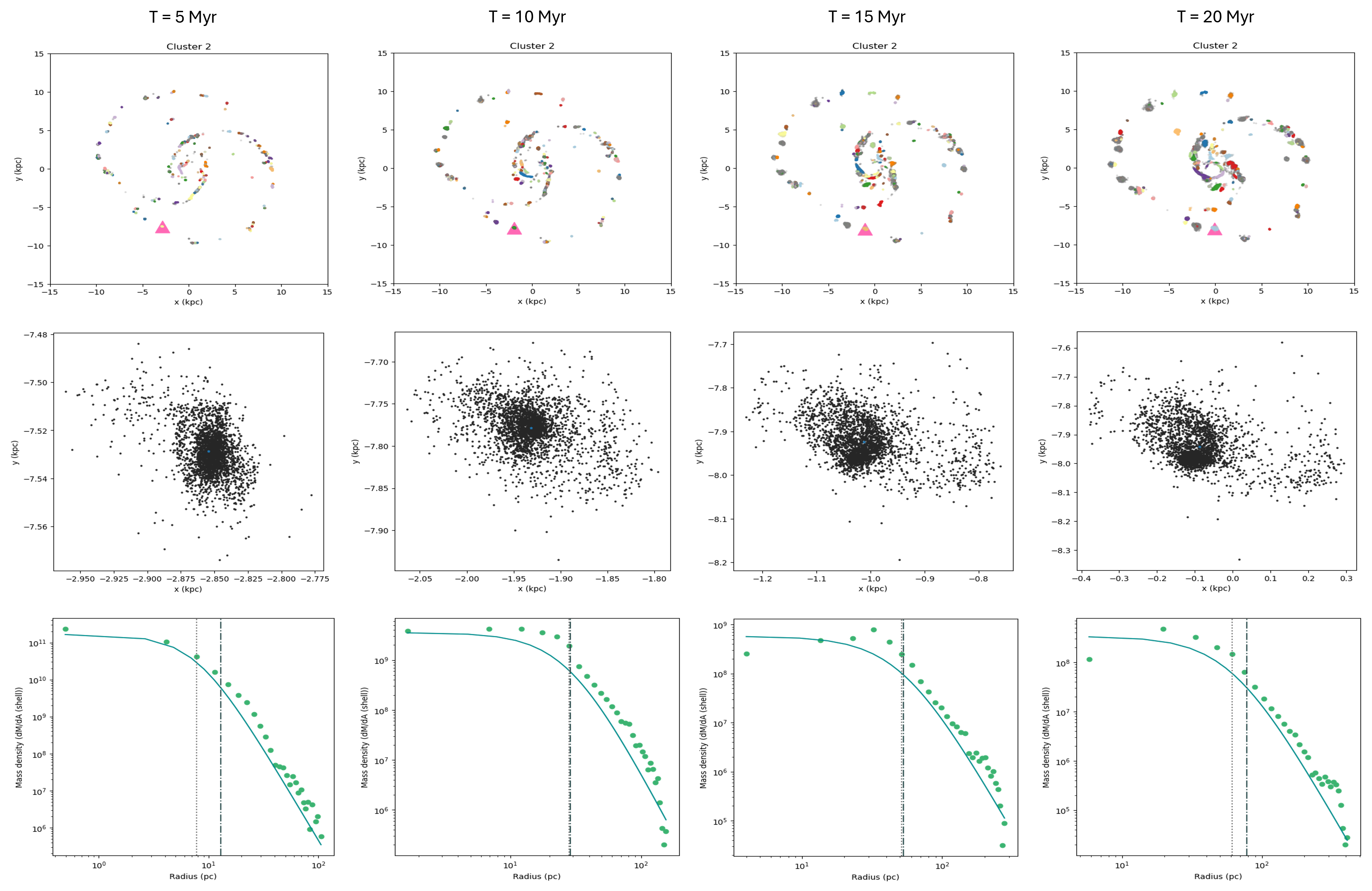}
    \caption{Cluster 2. Description of plots is the same as figure \ref{cluster_0_timeseries}.}
    \label{fig:cluster_2_timeseries}
\end{figure*}

\begin{figure*}
    \centering
    \includegraphics[width=6.5in, height=4.0in]{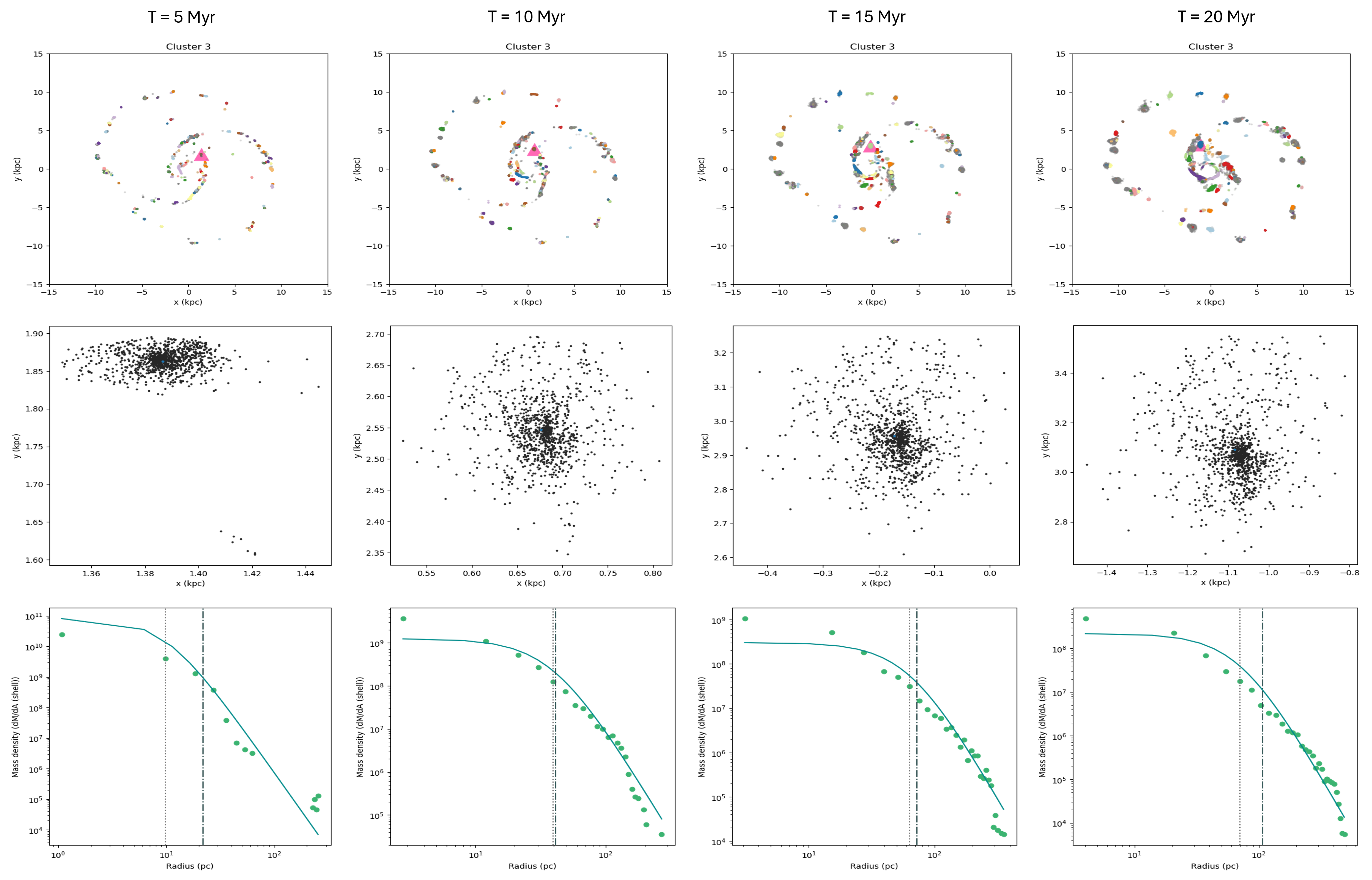}
    \caption{Cluster 3. Description of plots is the same as figure \ref{cluster_0_timeseries}.}
    \label{fig:cluster_3_timeseries}
\end{figure*}

\begin{figure*}
    \centering
    \includegraphics[width=6.5in, height=4.0in]{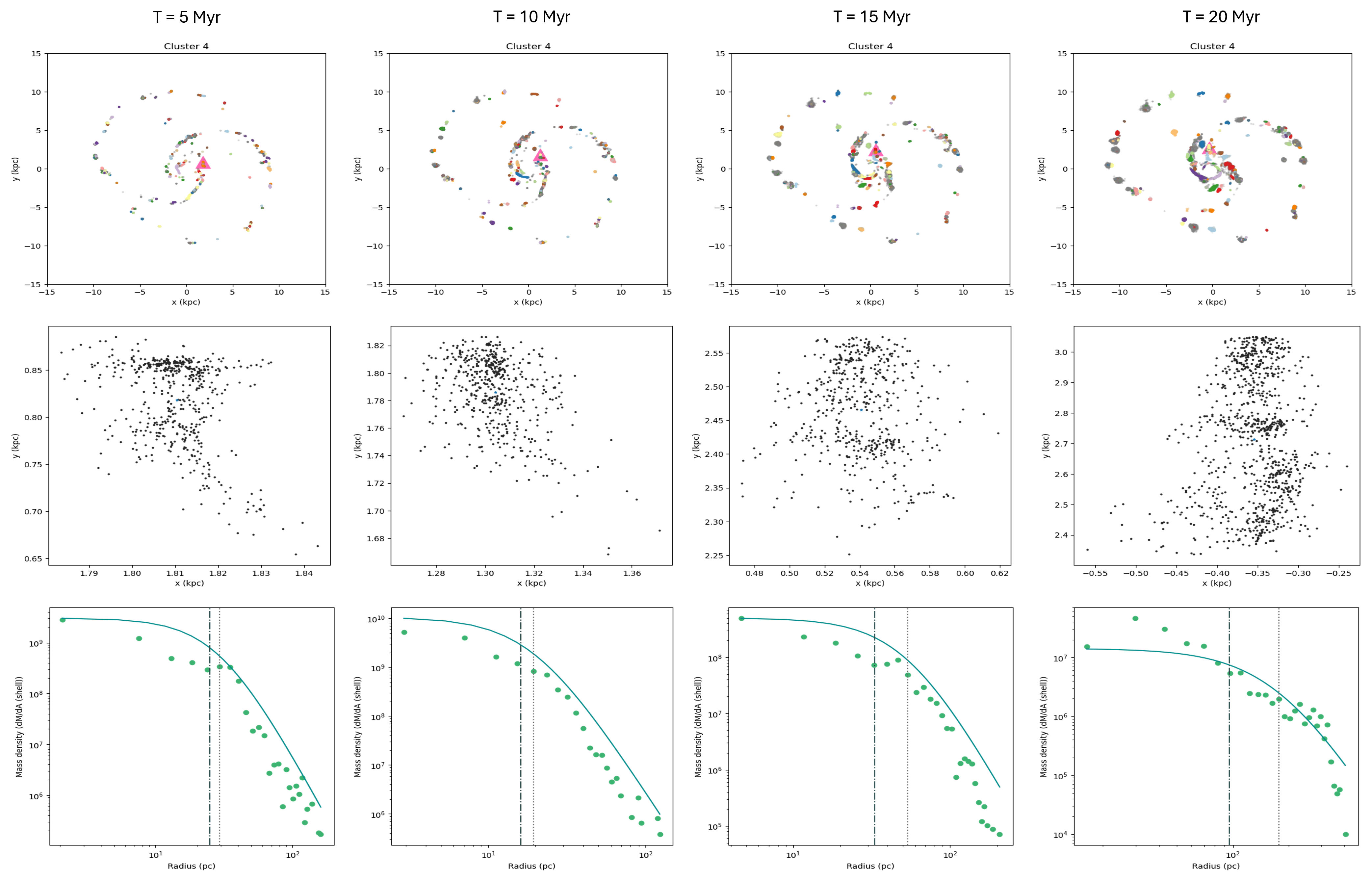}
    \caption{Cluster 4. Description of plots is the same as figure \ref{cluster_0_timeseries}.}
    \label{fig:cluster_4_timeseries}
\end{figure*}

\begin{figure*}[h]
    \centering
    \includegraphics[width=6.5in, height=4.0in]{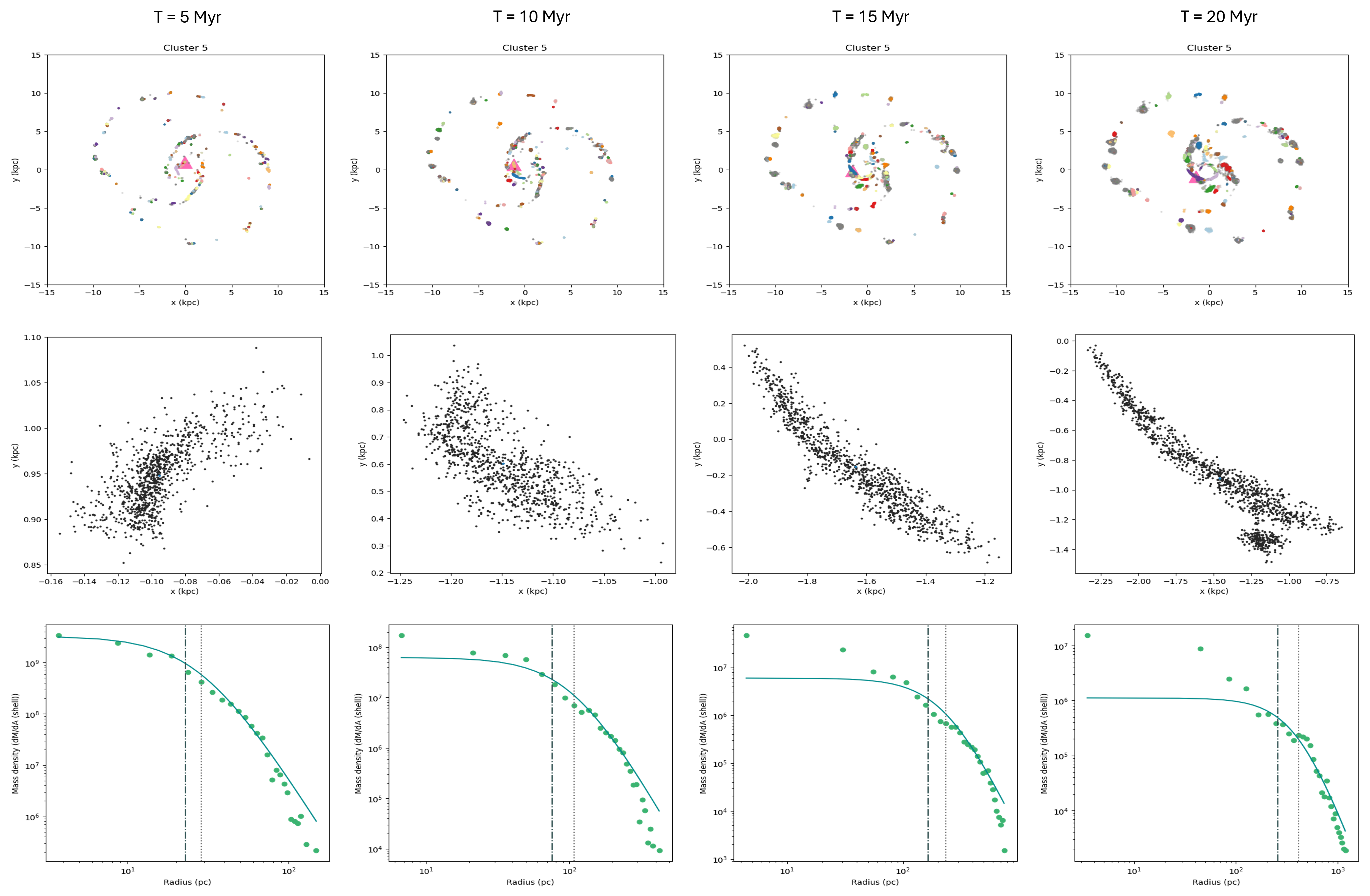}
    \caption{Cluster 5. Description of plots is the same as figure \ref{cluster_0_timeseries}.}
    \label{fig:cluster_5_timeseries}
\end{figure*}

Figures \ref{fig:cluster_1_timeseries}, \ref{fig:cluster_2_timeseries}, \ref{fig:cluster_3_timeseries}, \ref{fig:cluster_4_timeseries}, and \ref{fig:cluster_5_timeseries} show the time evolution over 20 Myr in steps of 5 Myr for five individual clusters. The first row shows the location (pink triangle) of the cluster in the galaxy. The second row shows the scatter plot of individual star particles in the cluster. The last row shows the mass density profile where the dotted line represents the half-mass radius and the dash-dotted line is the standard deviation radius. The solid blue line is the Plummer profile where we use the half-mass radius as the Plummer radius. The half-mass radius and mass values for each cluster at each time are in Table \ref{table:clust_evoln}.

\begin{table}
\caption{Cluster Time Evolution}
\centering
\begin{tblr}{
  @{}ccccccX[c,valign=b]X[c,valign=b]X[c,valign=b]@{}
}
\hline
\hline
 Cluster \#&
 Time (Myr) &
 Half-mass Radius (pc) &
 Mass ($M_\odot$) \\
\hline

\noindent\SetCell{3}{Cluster 0}& 5 & 8.72 & $2.75\times10^5$ \\ 
& 10 & 26.48 & $2.69\times10^5$ \\
& 15& 37.41 & $2.34\times10^5$ \\
& 20 & 16.55 & $9.12\times10^4$ \\
\hline

\SetCell{3}{Cluster 1}&5 & 12.77 & $9.12\times10^5$ \\
&10 & 43.91 & $9.12\times10^5$ \\
&15& 9.44 & $2.29\times10^5$ \\
&20 & 11.8 & $2.51\times10^5$ \\
\hline

\SetCell{3}{Cluster 2} & 5 & 7.76 & $9.12\times10^5$ \\
&10 & 28.06 & $8.91\times10^5$ \\
&15& 51.31 & $7.76\times10^5$ \\
&20 & 61.34 & $7.56\times10^5$ \\
\hline

\SetCell{3}{Cluster 3} &5 & 9.78 & $3.16\times10^5$ \\
&10 & 39.67 & $3.16\times10^5$ \\
&15& 63.46 & $2.88\times10^5$ \\
&20 & 70.58 & $2.88\times10^5$ \\
\hline

\SetCell{3}{Cluster 4} &5 & 29.42 & $1.62\times10^5$ \\
&10 & 19.44 & $1.62\times10^5$ \\
&15& 53.74 & $1.48\times10^5$ \\
&20 & 176.48 & $2.57\times10^5$ \\
\hline

\SetCell{3}{Cluster 5} &5 & 28.68 & $3.02\times10^5$ \\
&10 & 107.45 & $3.09\times10^5$ \\
&15& 234.9 & $2.82\times10^5$ \\
&20 & 411.60 & $3.31\times10^4$ \\
\hline

\end{tblr}
\label{table:clust_evoln} 
\end{table}

\section{HDBSCAN in Two Dimensions}\label{app_c}

\begin{figure*}[ht]
    \centering
    \includegraphics[width=7in, height=4in]{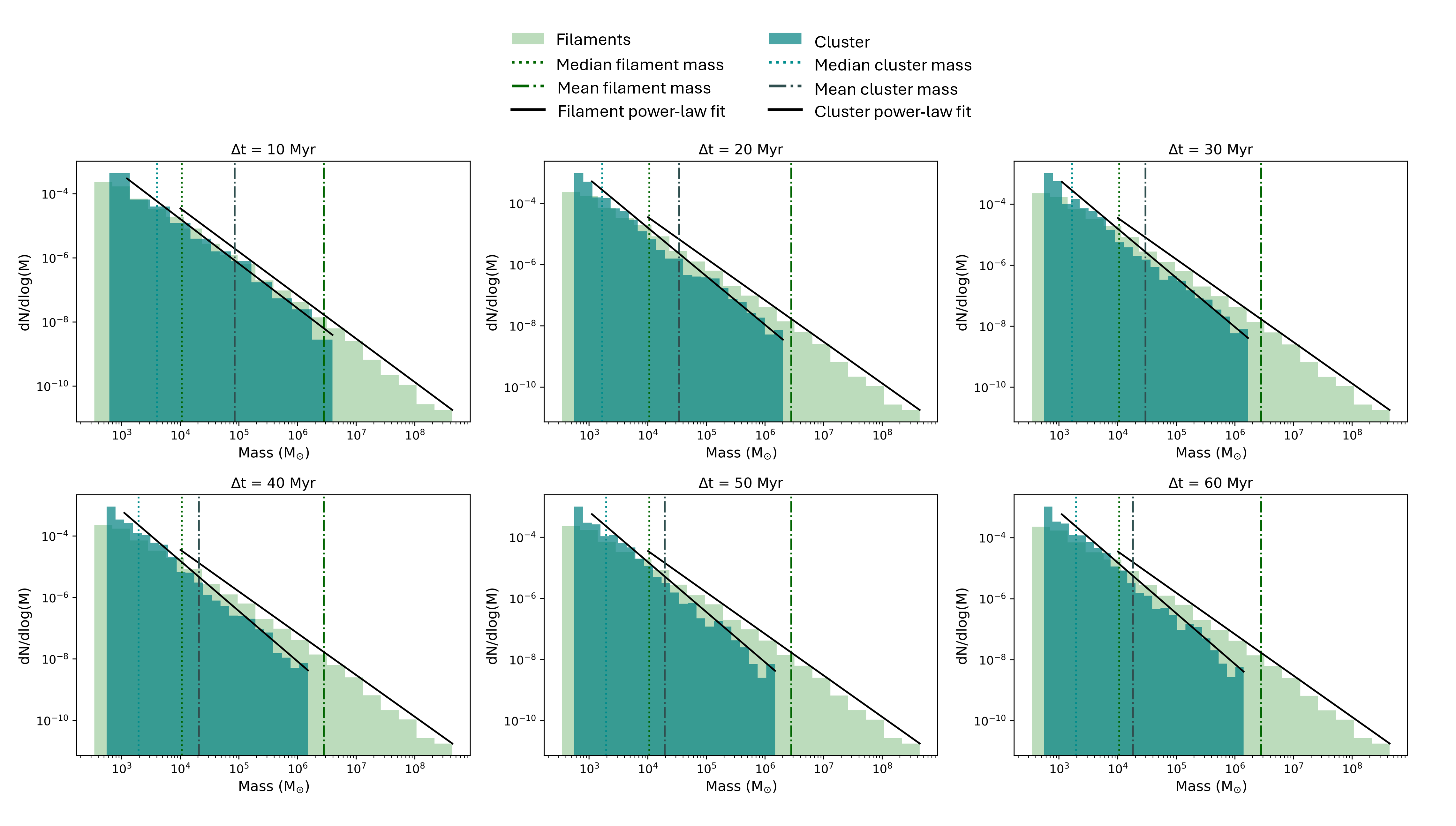}
    \caption{Six plots of the mass distribution of filaments (green) and clusters (teal) identified in 2D. The filament mass distribution is of all filaments at a time of 327 Myr. The cluster mass distribution is of the star particle population formed between 268-278 Myr and identified by applying HDBSCAN to the projected particle positions. Each frame evolves the star particle population by 10 Myr, ending at 327 Myr. The black lines show the power-law fit to the filament and cluster mass distribution. The dotted and dash-dotted lines represent the median and mean masses respectively, colour coded to identify the cluster or filament distribution. }
    \label{fig:cluster_histos_2D}
\end{figure*}

To determine if projection effects alter our results, we ran an experiment applying HDBSCAN to projected particle positions, rather than in three dimensions as described in \S \ref{sec:clustident}. We did this to ensure we can compare clusters to filaments identified from projected gas density. We found that our result that the power-law trend of clusters when they first form matches that of the filament mass distribution is still valid using HDBSCAN in 2D. The 2D power-law at $t = 10$ Myr is $-1.39$ using the powerlaw library compared to $-1.36$ in 3D. The trend of the mass distribution becoming steeper in 3D also persists in 2D as seen in figure \ref{fig:cluster_histos_2D} with slope values calculated via the powerlaw library in table \ref{table:2D_powerlaw}. This table shows both the 3D HDBSCAN power-law values from table \ref{table:clust_evoln} and the ones in 2D. We conclude that HDBSCAN in 2D does not meaningfully alter our results.

\begin{table}[ht]
\caption{2D vs 3D HDBSCAN Cluster Mass Power-laws}
\centering
\begin{tblr}{
  @{}cccccX[c,valign=b]X[c,valign=b]X[c,valign=b]@{}
}
\hline
\hline
 Evolution Time (Myr)&
 3D (powerlaw library)&
 3D (MCMC)&
 2D (powerlaw library)&
 2D (MCMC)\\
\hline
10 & 1.35 & $1.33^{+0.39}_{-0.41}$ & 1.39 & $1.38^{+0.42}_{-0.41}$\\
20 & 1.47 & $1.39^{+0.38}_{-0.38}$ & 1.58 & $1.48^{+0.31}_{-0.31}$\\
30 & 1.49 & $1.43^{+0.37}_{-0.38}$ & 1.61 & $1.51^{+0.31}_{-0.31}$\\
40 & 1.52 & $1.49^{+0.37}_{-0.37}$ & 1.63 & $1.59^{+0.33}_{-0.34}$\\
50 & 1.53 & $1.53^{+0.35}_{-0.36}$ & 1.64 & $1.64^{+0.36}_{-0.34}$\\
60 & 1.55 & $1.54^{+0.35}_{-0.36}$ & 1.66 & $1.65^{+0.35}_{-0.34}$\\
\hline
\end{tblr}
\label{table:2D_powerlaw}
\end{table}

\end{document}